\documentclass[12pt,aps,amsmath,latexsym,amsfonts]{JHEP3}

\usepackage{epsfig}
\usepackage{amsfonts,amssymb,amsmath}
\usepackage[hang]{subfigure}

\newcommand{\be}{\begin{equation}}
\newcommand{\ee}{\end{equation}}
\newcommand{\bea}{\begin{eqnarray}}
\newcommand{\eea}{\end{eqnarray}}

\def\Ab{{\bf A}}
\def\Bb{{\bf B}}
\def\ab{{\bf a}}
\def\bb{{\bf b}}

\def\r{{\bf r}}

\newcommand\vect[4]{\left(
\begin{matrix} #1 \\ #2 \\ #3\\ #4 \end{matrix}\right)}

\title{The effect of extra dimensions on gravity wave bursts from
cosmic string cusps}
\author{Eimear O'Callaghan$^a$\!
\thanks{Email: e.e.o'callaghan@durham.ac.uk}\ ,
Sarah Chadburn$^b$\!
\thanks{Email: s.e.chadburn@durham.ac.uk}\ ,
Ghazal Geshnizjani$^{c}$\!
\thanks{Email: ggeshnizjani@perimeterinstitute.ca}\ ,
\newline Ruth Gregory$^{a,b}$\!
\thanks{Email: r.a.w.gregory@durham.ac.uk}\ ,
Ivonne Zavala$^d$\!
\thanks{Email: zavala@th.physik.uni-bonn.de}\\
$~^a$ Institute for Particle Physics Phenomenology, Department of Physics,
Durham University, South Road, Durham, DH1 3LE, UK\\
$~^b$ Centre for Particle Theory, Department of Mathematical Sciences,
Durham University, South Road, Durham, DH1 3LE, UK\\
$~^c$  Perimeter Institute for Theoretical Physics,
31 Caroline Street North, Waterloo ON, N2L 2Y, Canada.\\
$~^d$ Bethe Center for Theoretical Physics and 
Physikalisches Institut der Universit\"at Bonn, 
Nu\ss allee 12, D-53115 Bonn, Germany.}

\abstract{
We explore the kinematical effect of having extra dimensions
on the gravitational wave emission from cosmic strings.
Additional dimensions both round off cusps, and 
reduce the probability of their formation.
We recompute the gravitational wave burst, taking into account
these two factors, and find a potentially significant damping
on the gravitational waves of the strings.
}

\preprint{DCPT-10/13, PI-cosmo-182}

\begin{document}

\section{Introduction}
\label{intro}

The notion that nature might have extra dimensions, and that these might
leave traces in observation or experiment, has led to many interesting
ideas and advances over the last decade. Traditionally, extra dimensions
would be hidden via a Kaluza-Klein mechanism, accessible only at ultra
high energies. More recently however, the braneworld 
paradigm, \cite{EBW,ADD,RS},
has allowed for extra dimensions to be much larger, hidden instead by
a confinement mechanism, \cite{EBW}, which localizes standard model physics
on the brane but allows gravity to probe the bulk. These Large Extra
Dimension (LED) scenarios, \cite{ADD,RS}, have the added attraction of
providing a natural hierarchy between gauge and gravity interactions
coming from geometric multiplying factors in the derived 4-dimensional
Planck mass. These ideas have been incorporated into string theory
models, allowing LED's via a process of flux stabilization, \cite{KKLT}.

One of the advantages of the braneworld scenarios was that they gave
concrete predictions for cosmology. In particular, the Randall-Sundrum
(RS), \cite{RS} set-up, with a brane living in anti de Sitter (adS)
spacetime with one extra dimension, has a simple cosmology,
the scale factor being 
determined by motion through a black hole spacetime, \cite{BCOS}.
This led to the mirage picture, \cite{mirage}, in which 
brane cosmology is determined by bulk motion through a warped space,
however, the mirage picture does not explicitly include the
gravitational back reaction of the braneworld, which for codimension
two and higher can be problematic, \cite{HCoD}. 

Nonetheless, the idea of the braneworld -- an object localized
in extra dimensions -- has proven to be extremely useful when
constructing models of inflation in string theory. The localization
in extra dimensions provides a natural scalar field (the position
of the brane) which can then evolve according to some effective
action depending on the specifics of the internal manifold and
the brane itself. Early models of brane inflation, \cite{braneinf},
used the Newtonian potential of the higher dimensional
solution to provide the inflaton potential. This however,
was not consistent with the size of the extra dimensions, nor
was any mechanism offered for their stabilization. However,
with the completion of the picture via flux stabilization,
\cite{KKLT,KKLMMT}, a great many models of brane inflation
have been explored (see \cite{BI} for a review).

A key side effect of brane inflation is the formation of cosmic
strings, \cite{BCS,JST}. These strings form as a by product of the 
annihilation of the inflationary branes (for reviews see \cite{CSSrev}), 
and while having their origins in superstring theory, can have a 
wide range of parameters, and interesting physical properties. 

Cosmic strings, \cite{CS}, are an example of a topological defect, a glitch
in the vacuum structure of a field theory which can arise when
the vacuum manifold is topologically nontrivial. They are ubiquitous
in all types of physics, from condensed matter systems to 
quantum field theory. They were first explored in the cosmological
setting by Zeldovich and Kibble, \cite{ZelKib}, who realised that 
they could arise in the early universe
as possible side-products of symmetry breaking phase transitions.
Indeed, it was the cosmological catastrophe of monopole production
at the GUT scale that in part led to the development of the original
inflationary scenario.
Cosmic strings however are cosmologically benign, and exhibit a
scaling network behaviour, \cite{NET}. Accounting for their gravitational
effects made them a possible candidate for the perturbation spectrum
until CMB experiments ruled them out, \cite{CMBrout}. 
Nonetheless, the possibility remains that a network of light 
strings could be present in our universe, \cite{CSfit}.

From the cosmological point of view, the internal structure of the
cosmic string is irrelevant, and what we need is the long range 
behaviour of the string. Gravitationally, a straight string produces
a conical deficit in spacetime \cite{cone}, which, while it does 
lead to interesting lensing effects, cannot be regarded as a detection
tool as it requires the serendipity of an appropriately aligned source
behind the string (although see \cite{KK} for a discussion of detection
via weak lensing). Instead, a more promising approach is to take
the dynamical network of strings, and to use linearized gravity
to compute the radiation emitted from the loops and crinkly long
strings \cite{gwave,GWL,Eloss}. Network simulations approximate the 
string by the Nambu action:
\be
\label{NGaction}
S = -\mu \int d^2 \sigma \sqrt{\gamma}
\ee
which can be rigorously derived from an underlying field
theory model \cite{NG}. Together with rules for intercommutation
\cite{IC}, or how strings behave when they cross each other, this
gives the basic physics of how a network evolves.

The primary drivers of network evolution are therefore whether 
strings intersect, and whether the motion has any extreme 
events (i.e.\ are there points at which the Nambu approximation 
is likely to break down). Early work focussed on taking toy 
families of loops, \cite{KT,loops}, to explore the likelihood of self
intersection, and also events where the string bends on the 
same scale as its width. The kinematics of strings obeying the
Nambu action are very well understood: the picture is that 
loops often self
intersect, and generically have cusps -- points at which the
string reaches the speed of light and its extrinsic
curvature diverges, leading to a breakdown of the 
Nambu action. Finite width corrections to the Nambu action can
be computed, \cite{FWC,EffAction}, but it was believed that these would not
be significant for the network evolution.
In appendix \ref{AppA} we demonstrate this explicitly by computing the 
finite width effect for a cusp.
More recently, further analysis of the effect of
small scale structure on the string, \cite{SSS}, has indicated that 
there may actually be measurable consequences.
In either case however, the primary importance of a cusp is that 
it acts as a strong source of gravitational radiation.

In a seminal paper, Damour and Vilenkin (DV), \cite{DV}, re-examined 
gravitational radiation from cosmic strings, assessing for what
range of mass per unit length the string could potentially be visible
to the next range of gravitational wave detectors. The main effect they
were considering was the burst of radiation from extreme kinematic
events in the loop motion, known as cusps and kinks.
They computed the amplitude of a cusp and kink gravitational wave
burst (GWB) as a function of the mass per unit length of the string. 
In a later paper, \cite{DV2}, they allowed for 
networks formed by strings with lower intercommutation probabilities, 
which enhance the density and thus the GWB amplitude.
Siemens et al.\ \cite{Siem,SiemStoch}, performed a more careful analysis of the 
cosmological expansion history, instead computing rates of events
at amplitudes fixed by the detector. Since then, many other gravitational
effects of cosmic superstrings have been explored, including 
strings with junctions, \cite{GWCSS}, and broken strings, \cite{beads}.

In a recent note \cite{CCGGZ}, we revisited the calculations of DV et al.\
arguing that the kinematic effect of the extra dimensions significantly
reduced the power of the cusp waveform. 
Essentially, extra dimensions act to `slow down'
the string, as first pointed out by Avgoustidis and Shellard, \cite{AS},
and round off the sharp cusp. In this paper we give a full computation
of this effect, presenting test loop families to demonstrate
parameter space measures, and detailed numerical calculations of
GWB event rates and amplitudes.

The extra dimensions give two main modifications to the DV result.
The first is the cusp rounding effect, and we show how this gives 
a high frequency cut-off to the gravitational waveform. The second
is a probabilistic factor: in $3+1$ dimensions, cusps always form on a
smooth loop trajectory, however, in higher dimensions this is no
longer the case. DV introduced a parameter in their GWB calculation,
${\cal C}$, which measured the probability of cusp formation in a
single loop period; the reason for introducing this in $3+1$ dimensions
is that once strings intersect, they can have kinks, which lower
the cusp probability. In higher dimensions however, the probability of 
cusp formation is strictly zero, in that the set of solutions which 
have cusps has measure zero. We therefore define a {\it near cusp 
event}, which represents the rounded cusp, and compute the probability 
of this event to input in the amplitude calculation. 

Finally, one has to factor in the reduction in intercommutation
probability due to the extra dimensions, \cite{ICSC,AS2}, 
which acts to increase the density of the network. Summing all these
effects produces a marked effect on the GWB amplitude, and 
our conclusion is that detection of GWB's from cosmic superstrings by
current or next generation gravitational wave detectors will be harder
than suspected, with the bonus that positive detection may tell us
something about the number of extra dimensions. 
Clearly, our result will also relax bounds on $G\mu$ for cosmic
superstrings derived using the DV results \cite{obsv}.

We start by reviewing the standard Kibble-Turok method of 
analysing string trajectories, \cite{KT}, noting the new features
that appear with additional dimensions. We then review the DV
calculation of the GWB. Next we calculate the
various effects coming from extra dimensions, introducing a
parameter $\Delta$ which measures deviation from an exact cusp.
We integrate over $\Delta$ to obtain the sum of all
cusp or near cusp events, and finally discuss
implications and caveats of the calculation.
To be specific, we focus on the frequency band of the
advanced LIGO detector, however we will comment on the
frequency dependence of our results.

\section{String motion and cusps}
\label{stringSetup}

We begin by briefly reviewing the kinematics of cosmic strings,
deriving the general form of a string solution and showing how
cusps are generic. This formulation was largely developed by
Kibble and Turok, \cite{KT}, and is the standard method for
finding loop trajectories.

Let $X^\mu(\sigma^A)$ be the spacetime coordinates of the string 
worldsheet, where $\sigma^A=\{ \tau, \sigma\}$ are intrinsic
coordinates on the worldsheet. For closed loops, which we will be 
considering in this paper, $\sigma \in [0,L]$, where $L$ 
is the length of the loop. The induced metric on the worldsheet 
appearing in (\ref{NGaction}) is then:
\be
\gamma_{AB}= \frac{\partial X^\mu}{\partial \sigma^A}
\frac{\partial X^\nu}{\partial \sigma^B}\, g_{\mu\nu}
\label{intrinsic}
\ee
where $g_{\mu\nu}$ is the spacetime metric\footnote{We use a mostly minus
signature.}. The Nambu-Goto action (\ref{NGaction}) 
is then proportional to the area of the string worldsheet,
with the constant of proportionality being
$\mu$, the tension, or mass per unit length of the string.
Note that cosmic strings have a tension along their length equal 
to their energy density.

Because we are dealing with a two-dimensional metric, we can always
choose a gauge in which $\gamma$ is conformally flat:
\bea
\label{confgauge}
\dot{X}^\mu X^{\prime\nu} g_{\mu\nu} =0\\
\left ( \dot{X}^\mu \dot{X}^\nu
+X^{\prime\mu} X^{\prime\nu} \right ) g_{\mu\nu} =0 \; ,
\eea
where a dot denotes $\partial/\partial\tau$ and 
a prime $\partial/\partial\sigma$.
Kibble and Turok then chose the spacetime coordinates to coincide
with the centre of mass frame of the string, and the worldsheet 
time coordinate to correspond with the spacetime time (temporal
gauge). Thus writing $X^\mu=(\tau, {\bf r}(\tau, \sigma))$, we
have:
\bea
\dot{\r}\cdot\r^\prime=0\label{gauge1}\\
\dot{\r}^2+\r^{\prime2}=1\label{gauge2}\\
\ddot{\r}-\r^{\prime\prime}=0 \label{KTeom}
\eea
where the first two correspond to the gauge constraints, and the final
equation is the wave equation of motion for the string. It is conventional
to use lightcone coordinates:
\be
\sigma_\pm = \tau\pm \sigma
\ee
in which the solutions to the equation of motion (\ref{KTeom}) take the
form of left and right moving waves, conventionally written in the form
\be
\r= \frac{1}{2}[\ab(\sigma_-)+\bb(\sigma_+)],
\ee
where the gauge conditions constrain $\ab'$ and $\bb'$ to lie on a unit
sphere, commonly dubbed the ``Kibble-Turok'' sphere:
\be \label{unitcond}
\ab^{\prime 2}=\bb^{\prime 2}= 1\;.
\ee
Notice that while the periodicity of $\ab$ and $\bb$ is $L$, the 
periodicity of the actual motion of the string is $L/2$, since 
$\r(\sigma+L/2,\tau+L/2)=\r(\sigma,\tau)$.

There is an additional constraint that must be satisfied by
both $\ab'$ and $\bb'$, for consistency with the facts that the loop
is closed, and that we are in the {\it c.o.m.}\ frame. The former
condition requires that $\r(\tau,0) = \r(\tau,L)$, hence 
\be
\int_0^L \r' d\sigma = \int_0^L \left ( \bb' - \ab' \right ) 
d\sigma = 0\;.
\ee
The latter condition requires the average momentum integrated along the
string to vanish, i.e.:
\be
\int_0^L \dot{\r}\, d\sigma = \int_0^L \left ( \bb' + \ab' \right ) 
d\sigma = 0 \;
\ee
thus
\be
\int_0^L  \bb' \, d\sigma  =  \int_0^L  \ab' \, d\sigma  = 0\;.
\ee
Hence $\ab'$ and $\bb'$ follow trajectories on a unit sphere with 
zero weight -- their average position is the origin. Since they
both define curves on a two dimensional manifold which must cover 
both halves of the sphere equally, they will in general cross.
Inserting the expression for $\r$ into the intrinsic metric
(\ref{intrinsic}) gives:
\be
\gamma_{AB}=
\frac{1}{2}(1-\ab'\ldotp\bb')\eta_{AB} \, ,
\label{gammaAB}
\ee
thus when $\ab'$ and $\bb'$ are collinear, the metric becomes degenerate
and a point of the worldsheet instantaneously reaches the speed of light.
Strictly speaking the mass concentration on the string is infinite at this
point, however as it has zero area the total energy is finite. However,
since the vicinity of this point is highly relativistic, this rapidly
moving part of the worldsheet will have high momentum, and hence we
expect some significant gravitational interaction. Cusps are thus
transient but powerful events; moreover, they are generic on string
trajectories (notwithstanding the effect of small scale structure \cite{SSS}).
We now turn to a summary of the gravitational effects of cusps.

\section{Gravitational waves from cusps}
\label{gravitywaves3d}

It is worth reviewing the Damour-Vilenkin argument, \cite{DV}, as the
derivation of the gravitational wave signal is quite involved and 
lengthy\footnote{Note, DV use the mostly plus metric convention, hence some
equations will have relative minus signs compared with those we present
here.}.  Damour and Vilenkin first
computed the linearized metric perturbation arising from a single cusp 
event on a cosmic string loop of length $L$ in flat spacetime. The 
waveform of the cusp was found to have a power law behaviour of $f^{-4/3}$
(or $f^{-1/3}$ in their logarithmic Fourier representation) at large
frequencies, $f$, of the gravitational wave. 
They then used this flat spacetime waveform to infer the
cosmological waveform behaviour in the geometric optics limit, thus 
deriving a gravitational wave amplitude of a single cusp event which decays
quite strongly with redshift. 
Finally, by considering a one scale model for the string network, they 
computed an event rate for observing cusp GWB's 
which increased rapidly with redshift $z$. 
By choosing a physically reasonable event rate, and
picking a fiducial experimentally motivated frequency, they determined
the typical redshift contributing to the GWB and calculated the amplitude
of the cusp signal, presenting the results as a function of $G\mu$.

In order to present an analytic argument, DV introduced various
interpolating functions in redshift space, and approximated at
various stages the exact expressions in the waveform. As we review
their argument, we will keep these exact expressions until the final
stage of the calculation. When adding in the effect of the extra
dimensions, we will first follow the same game as DV, introducing 
the same interpolating functions so that a direct comparison can
be made. For interest however, we also include an exact numerical 
redshift integration.

The first step is to calculate the gravitational wave of a cosmic string 
loop in flat spacetime. We therefore need to solve the linearized 
Einstein equations
\be
\Box {\bar h} _{\mu\nu} = -16\pi G T_{\mu\nu}
\ee
which in the far field approximation is given by
\be
{\bar h} _{\mu\nu} \simeq \frac{4G}{r} \sum_\omega
e^{-i\omega(t-r)} T_{\mu\nu} ({\bf k}, \omega)
\ee
where $T_{\mu\nu} ({\bf k}, \omega)$ is the Fourier transformed
energy momentum.

The energy momentum of the cosmic string is
\be
T^{\mu\nu} = \mu \int d^2\sigma ({\dot X}^\mu {\dot X}^\nu - X^{\prime\mu}
X^{\prime\nu}) \delta ^{(4)} (x^\mu - X^\mu(\sigma,\tau))
\ee
which means the gravitational wave is determined by the Fourier transform
\be
T^{\mu\nu} ({\bf k},\omega) = \frac{\mu}{T_L} \int_0^{T_L} d\tau
\int_0^L d\sigma {\dot X}^{(\mu}_+ {\dot X}^{\nu)}_-
e^{-\frac{i}{2}(k\cdot X_+ + k\cdot X_-)}
\ee
where $X_+^\mu = \left (\sigma_+, {\bf b} (\sigma_+) \right)$, 
$X^\mu_- = \left (\sigma_-, {\bf a} (\sigma_-) \right)$,
and a dot now denotes a derivative with respect
to the argument of $X_\pm^\mu$; $k^\mu = \frac{4\pi m}{L} (1, {\bf n}) 
= m \omega_L (1, {\bf n})$ is the null wave vector.
Here, $\omega_L$ is the frequency of the fundamental mode of the string loop.

A cusp corresponds to a lining up of the momenta of the left and
right moving modes on the string loop: 
${\dot X}^\mu_+ = {\dot X}^\mu_-=\ell^\mu = (1,{\bf n}')$.
Choosing the coordinate origins, we may write
\be
X^\mu_\pm (\sigma_\pm) = \ell^\mu \sigma_\pm + \frac{1}{2}
{\ddot X}^\mu _{_0\pm} \sigma_\pm^2 + 
\frac{1}{6} {\ddot X}\hskip -2mm{\dot{\phantom{O}}}\!^\mu_{_0\pm}
\sigma_\pm^3 \label{DVexpX}
\ee
where the subscript $0$ refers to evaluation at $\sigma_\pm=0$.
Now, defining the angle between $k^\mu$ and $\ell^\mu$ as $\theta$, which 
is assumed to be small, and writing $d^\mu = k^\mu-\ell^\mu = (0,{\bf d})$
(where $|{\bf d}| \simeq \theta$), and using the gauge conditions, we have:
\be
k_\mu X_-^\mu = m\omega_L \left [ \frac{1}{2} \theta^2 \, \sigma_- 
- \frac{1}{2}{\bf n} \cdot {\bf a}'' \, \sigma_-^2
+ \frac{1}{6} \left ( {\bf a}^{\prime\prime2} - {\bf d} \cdot {\bf a}'''
\right ) \sigma_-^3 \right ]
\label{DVkdotX}
\ee
together with a similar expression involving ${\bf b}$ and $\sigma_+$.
In the last bracket, the ${\bf d}\cdot \ab'''$ term is subdominant,
being of order ${\cal O}(\theta |\ab''|^2)$.

The two integrals in the energy momentum therefore take the form:
\be
I^\mu  = \int [ k^\mu - d^\mu + {\ddot X}^\mu \sigma ]
\exp \left [ - \frac{im\omega_L}{12} 
( 3 \theta^2 \sigma - 3 \theta |{\ddot X}| \sigma ^2
\cos\beta + |{\ddot X}|^2 \sigma^3 ) \right] d\sigma
\label{DVI}
\ee
where $\beta$ is the angle between $\bf d$ and $\ab''$.
As Damour and Vilenkin pointed out, the first $k^\mu$ term is a pure
gauge, however, when $d^\mu \neq 0$, it cannot be gauged away, as 
the product $k^\mu d_\mu \neq 0$, but the trace reversed
${\bar h}_{\mu\nu}$ must be tracefree. However, since correcting for
this simply introduces a subdominant term with the same
waveform as the main part of the perturbation, like DV, we simply
focus on the main part of the integral and compute the main 
contribution to the waveform. Thus, rewriting
\be
\label{IntReparam}
u = \left ( \frac{m \omega_L}{12} {\ddot X}^2 \right ) ^{1/3} \sigma
\qquad , \qquad \; \varepsilon = 
\left ( \frac{m \omega_L}{12 {\ddot X}} \right ) ^{1/3} \theta
\ee
the relevant part of the integral becomes
\be
\label{Iplusminus}
I = 
\left ( \frac{12}{m\omega_L {\ddot X}^2} \right ) ^{2/3} {\ddot X}
\int du (u-\varepsilon) \exp \left [ -i \left ( (u-\varepsilon)^3 
+ \varepsilon^3 + 3\varepsilon u^2 (1-\cos\beta) \right ) \right ].
\ee
For $\varepsilon \ll 1$, this integral is well approximated by the
$\varepsilon = 0$ value:
\be
I^\mu_\pm = - \left ( \frac{12}{m\omega_L {\ddot X}^2_\pm} \right ) ^{2/3}
\frac{i}{\sqrt{3}} \Gamma \left ( \frac{2}{3} \right ) {\ddot X}^\mu_\pm
\ee
and for $\varepsilon > 1$, the integral rapidly tends to zero
due to the oscillatory behaviour of the term proportional to
$(1-\cos\beta)$.
Thus DV obtain the logarithmic cusp waveform:
\be
h^{cusp}(f,\theta) \sim \frac{G\mu L^{2/3}}{r|f|^{1/3}} H[\theta_m - \theta]
\label{4Dwform}
\ee
where $H$ is the Heaviside step function, and $\theta_m$ is the 
critical value of $\theta$ for which the integral drops to zero:
\be 
\theta_m =  \left ( \frac{12{\ddot X}}{m\omega_L} \right ) ^{1/3}
\simeq \left (  \frac{2}{Lf} \right ) ^{1/3}
\label{thetamaxdef}
\ee
using ${\ddot X} \sim 2\pi / L$, and $f = m\omega_L/2\pi$.

To transform this to the cosmological setting, one essentially replaces
$f$ with  $(1+z)f$, where $z$ is the redshift at the time of 
emission of the cusp GWB, and we must replace
$r$ by the physical distance 
\be
a_0 r = a_0 \int_{t_e}^{t_0} \frac{dt}{a} = \int_0^z \frac{dz}{H}
= (1+z) D_A(z)
\ee
where $D_A(z)$ is the angular diameter distance at redshift $z$. 

Damour and Vilenkin next use the one scale model of a string network,
by writing 
\be 
L \sim \alpha t \qquad , \qquad \; n_L(t) \sim 1/ (\alpha t^3)
\label{onescale}
\ee
for the length and number density of the string network at
cosmological time $t$. Here $\alpha \sim \Gamma G\mu$ is a numerically
determined constant, \cite{gwave,GWL,Eloss}, presumed to represent the 
rate of energy loss from string loops via gravitational radiation.
As in DV, we will take $\Gamma \sim 50$, however see \cite{recentalpha}
for more recent work and discussion on this issue.

Finally, DV estimate the rate of GWB's observed around frequency $f$
coming from the spacetime volume in redshift interval $dz$:
\be
d{\dot N} \sim \frac{\nu(z)}{(1+z)} \frac{\pi \theta_m^2 (z)
D_A(z)^2}{(1+z)H(z)} dz
\label{dndot}
\ee
where the first factor of $(1+z)$ comes from the redshift of time
between emission and observation, $\nu(z)$ is the number of cusp
events per unit spacetime volume, and the final part is the measure
of the spacetime volume within the beaming cone at redshift $z$, where
the beaming cone angle at redshift $z$ is simply given by
\be
\theta_m(z) = \left ( \frac{2}{(1+z)fL(z)} \right )^{1/3}.
\label{thetaofz}
\ee
The number of cusp events is given by
\be
\nu \sim \,{\cal C}\, \frac{n_L}{P T_L} \sim \frac{2{\cal C}}{P\alpha^2 t^4}
\label{nudef}
\ee
where ${\cal C}$ is the average number of cusps per loop period
$T_L = L/2 \sim \alpha t/2$ and $P$ is the reconnection probability
of the strings. Classical strings which intersect almost always 
intercommute, thus $P=1$ \cite{IC}, however the existence of extra dimensions 
makes it easier for the strings to miss each other. This results in 
the reconnection probability $P$ being reduced, 
as strings which appear to meet in 3 dimensions could 
be missing each other in the extra dimensions, leading to 
an enhancement of the number density of loops in the string 
network \cite{DV2,JST}. More detailed simulations, however 
indicate that this result may be slightly modified \cite{AS2}. 

The final step of the DV argument is to integrate (\ref{dndot}) 
to find the rate 
\be 
{\dot N} = \int_0^{z_*} \frac{d{\dot N}}{d\ln z} d \ln z
\sim \frac{d{\dot N} (z_*)}{d\ln z}
\ee
and then substituting in a fiducial frequency and desired rate to
find the redshift which dominates the signal. Evaluating the
gravitational wave at this redshift and frequency then gives the 
amplitude.

We now revisit this argument with the addition of the effects
of the internal extra dimensions.

\section{Wave form in extra dimensions}
\label{gravitywavesextradim}

In computing the waveform with extra dimensions, there are several
features we need to consider \cite{CCGGZ}. 
First, there is the motion of the string in the extra dimensions,
as pointed out by Avgoustidis and Shellard \cite{AS},
which causes the strings to 
appear to slow down in our noncompact space dimensions. 
Next, there is the impact of this motion on the formation 
of cusps: as we will see, the effect of extra degrees of
freedom allows the left and right moving modes to misalign in
momentum space, thus avoiding an exact cusp in a similar way to
avoiding intercommutation. Finally, there is the gravitational
aspect of the extra dimensions. Since these strings are
formed in brane inflation scenarios, we will assume that the flux
stabilization procedure that prevents dangerous cosmological
moduli evolution also prevents the strings from exciting internal
degrees of freedom. Thus, we can use the normal 4D gravitational
propagator in calculating the gravitational radiation from a cusp.

\subsection{String kinematics with extra dimensions}

We begin with an overview of string solutions in $4+n$ dimensions.
As with 4 dimensions, these can be expressed in the Kibble-Turok
notation 
\be
{\bf R}= \frac{1}{2}[\Ab(\sigma_-)+\Bb(\sigma_+)],
\ee
where we use upper case to denote the full $3+n$-dimensional spatial
vectors, and lower case the noncompact dimensions. As before, 
$|\Ab'|^2 = |\Bb'|^2 = 1$, hence $\Ab'$ and $\Bb'$ trace out closed
curves on a unit $S^{2+n}$. 
Unlike in 3 space dimensions, where two
curves on an $S^2$ will generically cross, these curves will generically
miss each other. This means that the probability of an exact cusp
with extra dimensions is precisely zero. 
However, from the calculation of the GWB waveform, it is clear that
the power is radiated not exclusively from the cusp, but from a region 
in which the extrinsic curvature of the worldsheet is significant
(we will see shortly how the beaming cone opening angle defines this).
\FIGURE{\label{fig:nearcusp}
\includegraphics[width=7cm]{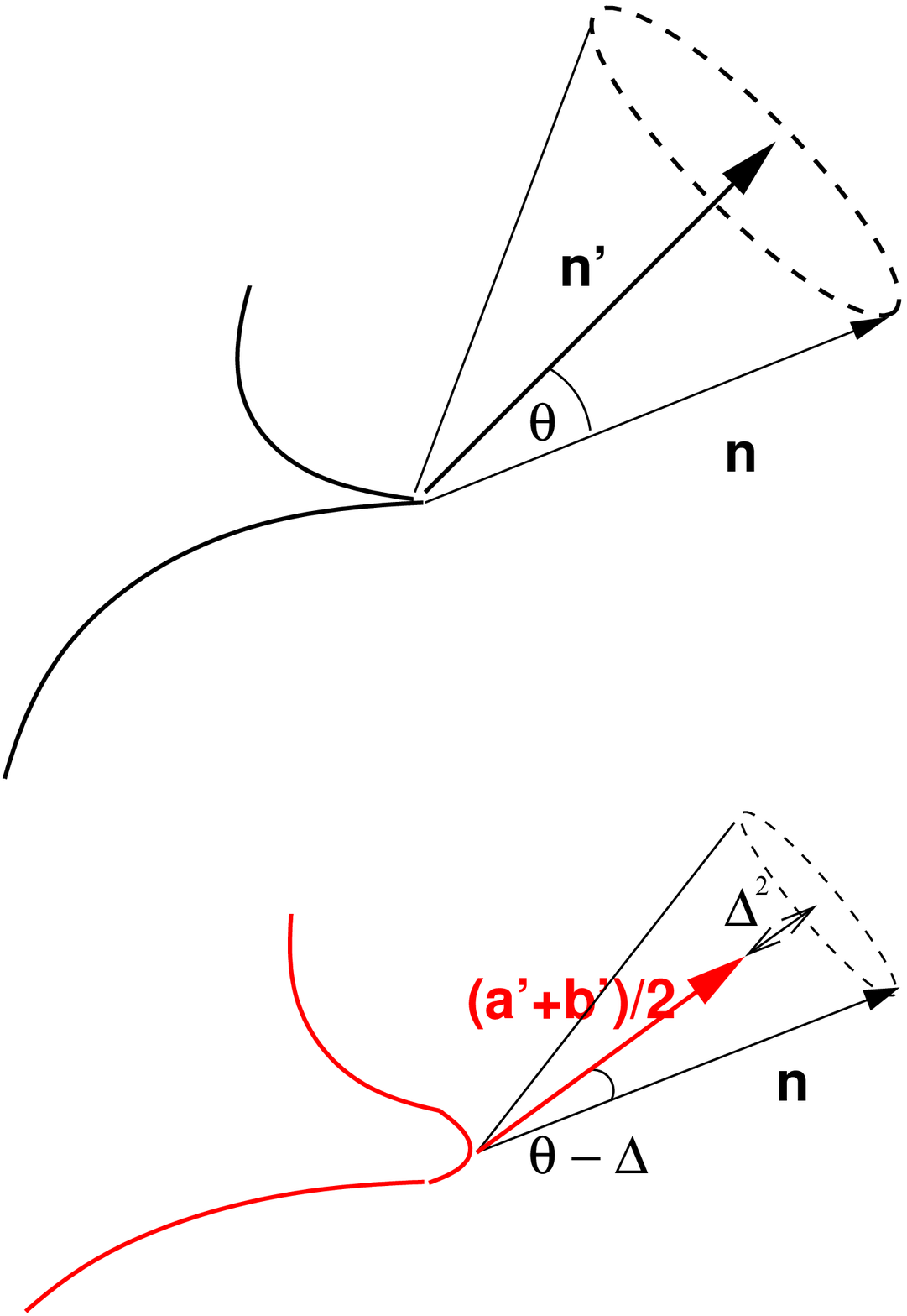}
\caption{A sketch of a near cusp event as opposed to an exact
cusp. The rounding of the cusp is indicated, as is the narrowing
of the beaming cone.
} }

We therefore generalise the exact cusp to a ``near cusp event'' (NCE)
for which
\be
|{\bf A}' - {\bf B}'| = 2\Delta
\label{NCdefn}
\ee
where $\Delta \ll 1$ is a parameter measuring how close to an exact cusp
(EC) we are. We can visualise the near cusp event as a rounded cusp,
as indicated in figure \ref{fig:nearcusp}.

In order to estimate the probability of near cusp formation, we
first assume an even measure in parameter space (we will discuss
alternative possibilities later). Each loop carries left and
right moving waves of harmonics of the fundamental frequency mode
$2\pi /L$, the wave vectors of which are constrained by the higher
dimensional version of the
gauge restriction (\ref{unitcond}). These can be represented in terms
of the rotation group $SO(n+3)$ \cite{Delaney}, and thus the
parameter space of the loop is simply parametrized by a set of angles.
An example of some low harmonic loops with one periodic extra 
dimension analogous to those considered
by Kibble and Turok \cite{KT} is given presently. These show
how the compactification of the extra dimension makes little difference to the self-intersection
probability for a zero width string, and demonstrates nicely the cusp
rounding effect.

We estimate the probability of NCE's therefore 
as $g(\Delta)^q$, where $q$ is the codimension in parameter
space of the subspace formed by loops which contain exact cusps,
and $g$ is a function which relates a shift in a parameter to a
change in $|\Ab' - \Bb'|$. $q$ 
can be readily computed from the condition for a cusp:
\be 
{\bf A}' = {\bf B}'\;.
\ee
This is a set of $n+3$ equations, however, as $|{\bf A}'| = |{\bf B}'|=1$
this results in $n+2$ constraints. Of these, two are used to fix
the values of $\sigma_\pm$ at the cusp, hence $n$ constraints
in parameter space remain. Thus, the codimension of the exact cusp
space is precisely the number of extra dimensions $q=n$.
In order to determine $g$, we modelled explicit loop solutions with
one extra dimension, and found that $g(\Delta) = g_0\Delta$, where
$g_0 \simeq 1$: see figure \ref{fig:cusp51}, where we plot
$\Delta=$ min$|\Ab' - \Bb'|/2$ against loop parameters.

\FIGURE{\label{fig:cusp51}
\includegraphics[width=8cm]{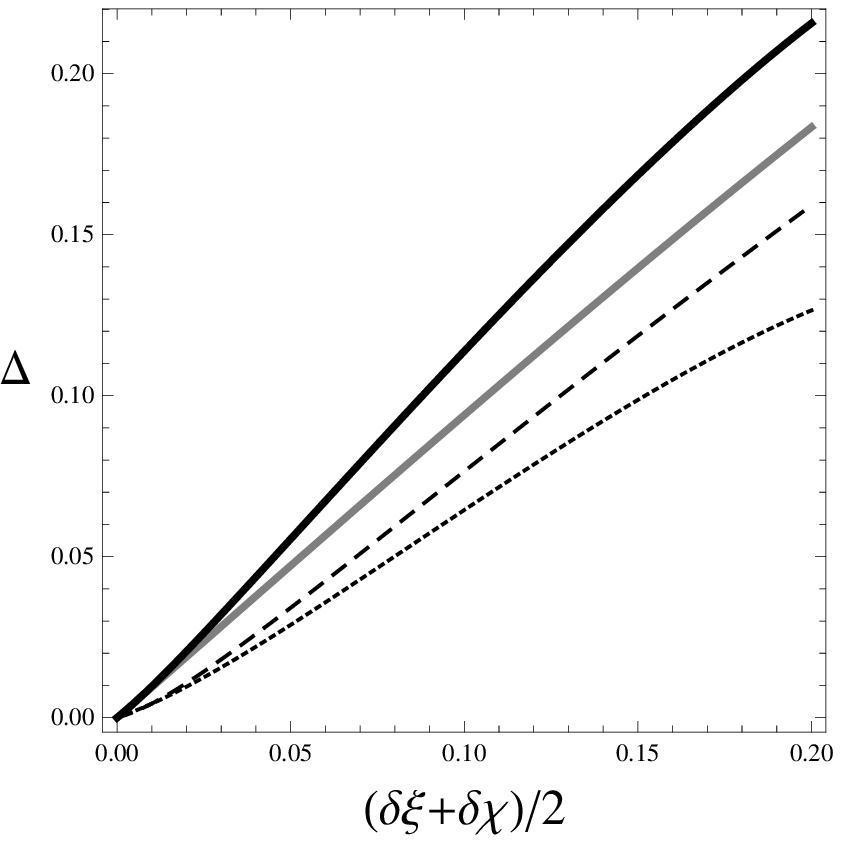}\nobreak\hskip 1cm
\includegraphics[width=6cm]{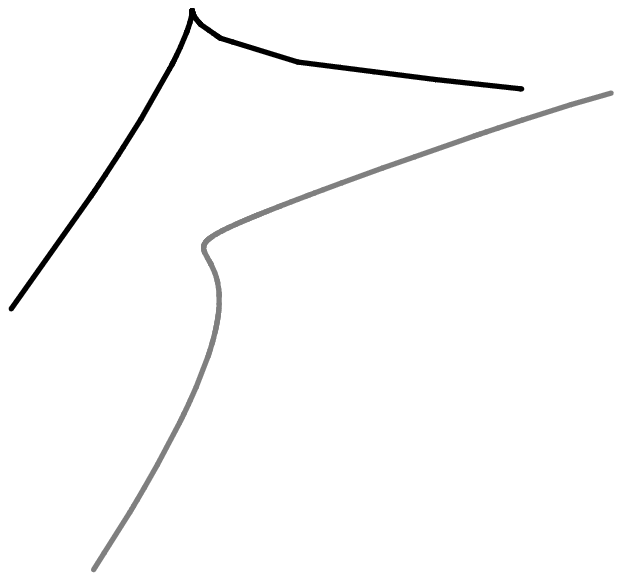}
\caption{
An examination of the dependence of the near cusp parameter, $\Delta$, 
on the loop solution parameters for the 1-5/1 loop family given in
(\ref{lvecs},\ref{rvecs}). The magnitude of $|\Ab'-\Bb'|$ is
computed as a function of the loop parameters as we move away
from an exact cusp event in parameter space. The `normal' direction
in solution space was computed using an expansion around the exact
cusp. The approximate linearity of the relation is demonstrated for
a range of parameter values and initial cusp values. An exact cusp occurs
when $\xi=-\chi$, $\zeta = \phi_1 = -\phi_2$. The black solid line 
corresponds to $\xi = \pi/4$, $\zeta = \pi/6$, the grey line to
$\xi = \pi/3$, $\zeta = \pi/4$, the dashed line to 
$\xi = \pi/4$, $\zeta = \pi/5$, and the dotted line to 
$\xi = \pi/4$, $\zeta = \pi/12$. The plot on the right shows the
effect of moving away from the cusp of the black parameter family
by a parameter shift of $0.1$.
} }

The outcome of our analysis is therefore that the number of NCE's
with $|\Ab' - \Bb'|_{\rm min} \leq 2\Delta$ in a generic loop is ${\cal N}
(\Delta) \simeq  \Delta^n$ (since all loops have $|\Ab' - \Bb'| \leq 2$ at all
points on their trajectory).
This argument of course simply refers to the cusps in the full 
higher dimensional loop motion, and not those loops which are close
to our 3-dimensional loops; it also makes no reference to any warping
of any of the spacetime dimensions. In addition, it assumes an
exact Nambu description, i.e.\ an exactly zero width string. The
strings will in general have finite width, and we expect that should 
the string width become a significant fraction of the internal dimension
size, then the motion in the internal dimension will be irrelevant.
Note however, that because these strings are basically classical
objects, there is no quantization of the motion in the internal directions.

A nice example of the effects of extra dimensions is given
by constructing a loop family. The general solution
for the left moving half for example is given by:
\be
\Ab (\sigma_-) = \sum_n \frac{L}{2\pi n} \, {\bf C}_n \,
\sin \left ( \frac{2\pi n\sigma_-}{L}\right )
+ \frac{L}{2\pi n} \, {\bf D}_n \, \cos \left ( \frac{2\pi n\sigma_-}{L}\right )
\ee
where the gauge conditions imply
\bea
2 &=& 
\sum_{n,m} \left ( {\bf C}_n . {\bf C}_m - {\bf D}_n . {\bf D}_m \right)
\cos \frac{2\pi (n+m)\sigma_-}{L}
+ \left ( {\bf C}_n  . {\bf C}_m + {\bf D}_n . {\bf D}_m \right)
\cos \frac{2\pi (n-m)\sigma_-}{L}
\nonumber \\
&+& \left ( {\bf C}_n  . {\bf D}_m + {\bf D}_n . {\bf C}_m \right)
\sin \frac{2\pi (n+m)\sigma_-}{L}
+ \left ( {\bf C}_m  . {\bf D}_n - {\bf C}_n . {\bf D}_m \right)
\sin \frac{2\pi (n-m)\sigma_-}{L}
\label{product}
\eea
A simple example of a new solution when we have one extra
dimension is to choose two independent harmonics $n>m$, with $n\neq3m$
(so that $2n, n+m, n-m,$ and $2m$ are all distinct).
The constraints from (\ref{product}) give 
\bea
{\bf C}_n . {\bf D}_n = 
{\bf C}_m . {\bf D}_m = 
{\bf C}_n . {\bf C}_m = 
{\bf D}_n . {\bf D}_m &=& 
{\bf C}_n . {\bf D}_m = 
{\bf C}_m . {\bf D}_n = 0\nonumber \\
{\bf C}_n ^2 = {\bf D}_n^2 \qquad ; \qquad &&
{\bf C}_m ^2 = {\bf D}_m^2 \\
{\bf C}_n ^2 + {\bf D}_n^2 +
{\bf C}_m ^2 + {\bf D}_m^2 
&=&2\;.\nonumber 
\eea
Thus we can take
\be
{\bf C}_m = \cos \zeta\ {\bf e}_1 \;,\;\;\;
{\bf C}_n = \sin \zeta\ {\bf e}_2 \;,\;\;\;
{\bf D}_m = \cos \zeta\ {\bf e}_3 \;,\;\;\;
{\bf D}_n = \sin \zeta\ {\bf e}_4 \;.
\label{lvecs}
\ee
Clearly this solution, with its requirement of 4 mutually orthogonal
vectors, is a simple example of a new solution in higher dimensions.
In three space dimensions (3D), we can only have a single harmonic,
unless $n=3m$. Thus the 3D limit of this left moving half is
$\zeta = 0,\pi/2$.
To give an illustrative loop family, we will take the right moving half
to have a single harmonic only
\be
\Bb (\sigma_+) = \frac{L}{2\pi} {\bf v}_1
\sin \left ( \frac{2\pi \sigma_+}{L}\right )
+ \frac{L}{2\pi} \, {\bf v}_2 \, \cos \left ( \frac{2\pi \sigma_+}{L}\right )
\ee
where ${\bf v}_1$ and ${\bf v}_2$ are two mutually orthogonal
vectors, which will be given by an SO(4) rotation of the $(x,y)$
plane: 
\be
{\bf v}_1 =
\vect{\cos\xi\cos\chi\cos\phi_1 - \sin\xi\sin\chi\cos\phi_2}
{\cos\xi\cos\chi\sin\phi_1 + \sin\xi\sin\chi\sin\phi_2}
{\sin\xi\cos\chi\cos(\phi_1+\phi_2) + \cos\xi\sin\chi}
{-\sin\xi\cos\chi\sin(\phi_1+\phi_2)}
\,;\;
{\bf v}_2 =
\vect{ \sin\xi\sin\chi\sin\phi_2 -\cos\xi\cos\chi\sin\phi_1 }
{\cos\xi\cos\chi\cos\phi_1 + \sin\xi\sin\chi\cos\phi_2}
{-\sin\xi\cos\chi\sin(\phi_1+\phi_2)}
{ \cos\xi\sin\chi -\sin\xi\cos\chi\cos(\phi_1+\phi_2) }.
\label{rvecs}
\ee
This loop family corresponds to a $m-n/1$ string in the notation
of \cite{loops}.

A time sequence of an evolving loop is shown in figure \ref{fig:Loop},
where we have taken $m=1$, $n=5$ to be specific, and set $L=2\pi$ for
convenience. A generic solution is compared with the 3D solution
with only one harmonic.
\FIGURE{\label{fig:Loop}
\includegraphics[width=10cm]{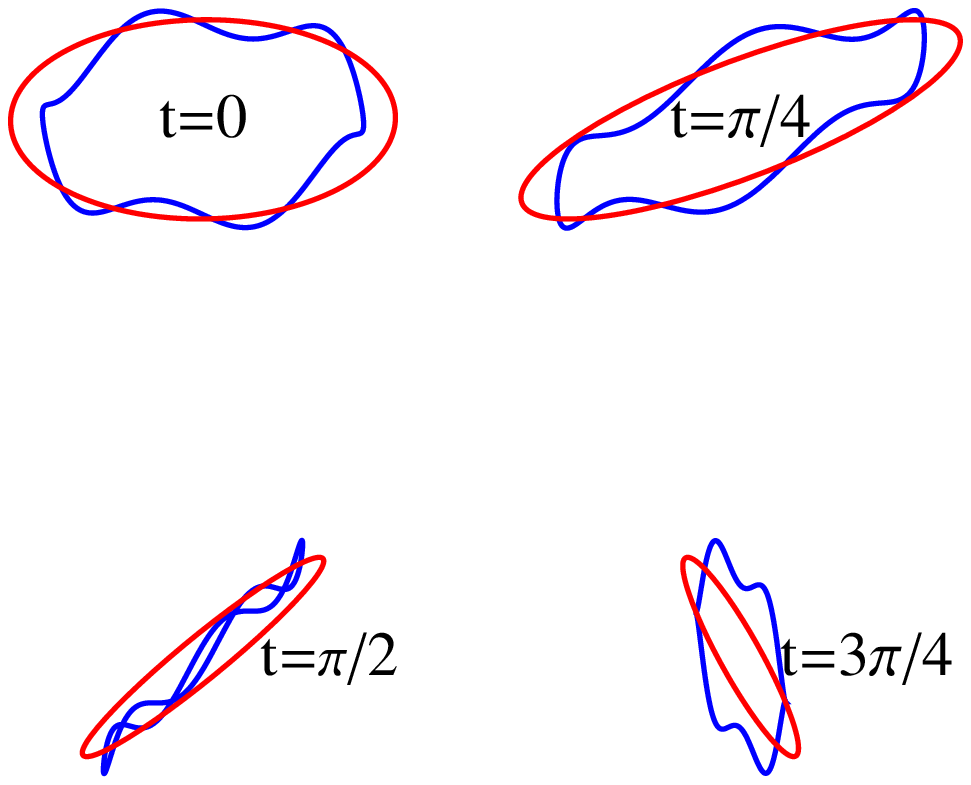}
\caption{
The time evolution of a 1-5/1 harmonic string with generic parameters
$\xi=-\chi=\pi/3$, $\phi_1=-\phi_2=\pi/4$ and $\zeta=\pi/4$ in blue, 
and the 3D loop with $\zeta = 0$ in red.
} }

\subsection{The gravitational waveform}

We now compute the waveform for a NCE with parameter $\Delta$.
The main difference between the EC and the NCE is that the 
velocity ${\dot X}^\mu = (1, (\ab'+\bb')/2)$ is now no longer 
necessarily null, and that the individual left and right moving
velocities need not be aligned. In other words, in evaluating the
integral (\ref{DVI}), we no longer have ``$\varepsilon=0$'', since
there are additional phase terms coming from the misalignment of
$\ab'$ and $\bb'$, as well as from the fact that ${\dot X}^\mu_\pm$ is
no longer null.

Define 
\bea
{\mbox{\boldmath$\delta$}}
&=& \frac{1}{2} ( {\bf a}'-{\bf b}' )\\
{\bf n}' &=& \frac{( {\bf a}'+{\bf b}' )}{|{\bf a}' +{\bf b}'|}
\eea
to be the separation vector of ${\bf a}'$ and ${\bf b}'$ at the NCE, and
the direction vector of the NCE respectively. 
Then writing 
$\Ab' = (\ab', {\mbox{\textbf{\textit{a}}}})$,
$\Bb' = (\bb', {\mbox{\textbf{\textit{b}}}})$, shows that
$|\ab'|^2 = 1-a^2$, and $|\bb'|^2 = 1-b^2$ (with $a
=|{\mbox{\textbf{\textit{a}}}}|$ etc.). A quick
check of (\ref{DVexpX}), (\ref{DVkdotX}) then indicates 
that the gravitational integral
(\ref{DVI}) will be damped unless $a, b \ll 1$.
While our modelling with compact extra dimensions indicates no
particular restrictions on the parameter space, the expectation
is that either warping of extra dimensions, or some other kinematic
consequence of cosmological expansion, will lead to the trajectories
being somehow close to the four dimensional behaviour (although
\cite{AA} indicates this may not be the case). We will therefore take
$a, b \ll 1$ from now on. Under these assumptions, expansion
of ${\mbox{\boldmath$\delta$}}$ and ${\bf n}'$ gives generically 
that $a^2 \sim b^2 = {\cal O} (\Delta^2) = {\cal O}(\delta)$.
Thus in orders of magnitude
\bea
{\bf a}' &=& \frac{1}{2} |{\bf a}' +{\bf b}'| {\bf n}' 
+ {\mbox{\boldmath$\delta$}} \simeq \left (1 - \frac{\Delta^2}{2} \right ) 
{\bf n}'  + {\mbox{\boldmath$\delta$}} \\
{\bf b}' &=& \frac{1}{2} |{\bf a}' +{\bf b}'| {\bf n}' 
- {\mbox{\boldmath$\delta$}} \simeq \left (1 - \frac{\Delta^2}{2} \right ) 
{\bf n}'  - {\mbox{\boldmath$\delta$}}\;.
\eea
Finally, estimating ${\bf n}' . \ab'' \sim {\bf n}' . \bb'' 
= {\cal O} (\Delta) |{\ddot X}|$
we find (making the same approximations as DV) the expression
\be
k_\mu X_-^\mu = \frac{1}{2} ( \theta^2 + \Delta^2) \, \sigma_- 
- \frac{1}{2}\left ( \theta + \Delta \right ) |{\ddot X}| \, \sigma_-^2
+ \frac{1}{6} {\ddot X }^2 \, \sigma_-^3
\ee
with a similar expression involving $X_+$ and $\sigma_+$.

Thus we find that the waveform
of the NCE is the same as that of the EC, with the proviso that the
cone opening angle is now decreased to
\be
\theta_\Delta = \theta_m - \Delta
\simeq \left (  \frac{2}{Lf} \right ) ^{1/3} - \Delta
\label{thetagendef}
\ee
i.e.\ the (logarithmic) NCE waveform is
\be
h^{\rm NCE} \sim \frac{G\mu L^{2/3}}{r|f|^{1/3}}
H[\theta_\Delta - \theta]\;.
\label{nDwform}
\ee
Notice that (\ref{thetagendef}) provides a high frequency cutoff to the
waveform, 
\be
f_\Delta = 1/(\Delta^3 T_L)
\ee
therefore our long frequency `tail' to the waveform is curtailed at
some (high) $\Delta$-dependent frequency. However, what is more relevant
cosmologically is the combination of the impact of this cutoff of the 
beaming cone area and the effect of the lowering of the number of NCE's.

Cosmologically, we need to calculate the GWB event rate ${\dot N}$ for
near cusp events, however, for a general network there will be a 
range of NCE's with different $\Delta$ values, up to and including 
the cutoff value when the GWB beaming cone closes off. 
We clearly need to integrate over these options to obtain the nett
effect of {\it all} possible NCE's. We therefore write
\be
\frac{d^{2}{\dot N}_{\rm NCE}}{dz\,d\Delta} \sim 
\frac{{\cal C}(\Delta) n_L(z)}{P T_L(z)} 
\frac{\pi \left (\theta_m(z) - \Delta \right )^2 D_A(z)^2}{(1+z)^2 H(z)}
\label{Ndelta}
\ee
where ${\cal C}(\Delta)$ is the local probability density of NCE's
for the network.  In four spacetime dimensions, a loop with 
continuous momentum functions always has a cusp, which would 
correspond to ${\cal C}(\Delta) = \delta(\Delta)$ in an
integration of (\ref{Ndelta}) (where $\delta(\Delta)$ is now the Dirac
$\delta$-function!). For extra dimensions, 
assuming that the loops are spread evenly in the parameter space 
of solutions, we get 
\be
{\cal C}(\Delta) = {\cal N}'(\Delta) = n\Delta^{n-1}
\ee
and hence the $\Delta$ integral yields
\be
\int_0^{\theta_m} {\cal C}(\Delta) 
\left ( \theta_m(z) - \Delta \right )^2
= \frac{2 \theta_m(z)^{n+2}}{(n+1)(n+2)}
\label{deltaint}
\ee
where the integral is saturated by $\theta_m<1$. Note that for the fiducial
frequency $f \sim 150$Hz, $\theta_m \sim 10^{-4} \to  10^{-2}$ 
as $G\mu \sim 10^{-6} \to 10^{-12}$ respectively, and since $\theta_m$ 
varies as $(1+z)^{1/6}$, it remains small until extremely high
redshifts $\left ( (1+z_{\rm rec})^{1/6} \simeq 3\right )$.

Gathering together these different effects, we therefore arrive at 
the expression for the GWB rate:
\be
\frac{d{\dot N}_{\rm NCE}}{dz} 
= \frac{2 \theta_m(z)^{n+2}}{(n+1)(n+2)}
\frac{n_L(z)}{PT_L(z)} \frac{\pi D_A(z)^2}{(1+z)^2 H(z)} \;.
\label{newrate}
\ee

Figure \ref{fig:All} shows the gravitational wave amplitude for the cosmic
string cusp bursts in the form presented by DV, \cite{DV2}, for
varying values of $n$. The graphs are presented first by calculating 
the amplitude in exactly the way DV did, by using interpolating functions,
and also neglecting $\Omega_\Lambda$. However, the dotted data curves
also show an exact redshift integration, keeping the precise values
of the angular diameter and cosmological time for the concordance
cosmology ($\Omega_r = 4.6 \times 10^{-5},\, \Omega_m = 0.28,\, \Omega_\Lambda
=1-\Omega_m-\Omega_r$) and integrating
out numerically in redshift space for different values of $G\mu$.
A similar plot is obtained for the LISA frequency band,
\cite{CCGGZ}, however, the suppression of the signal 
is less marked at lower frequency.
\FIGURE{\label{fig:All}
\includegraphics[width=14cm]{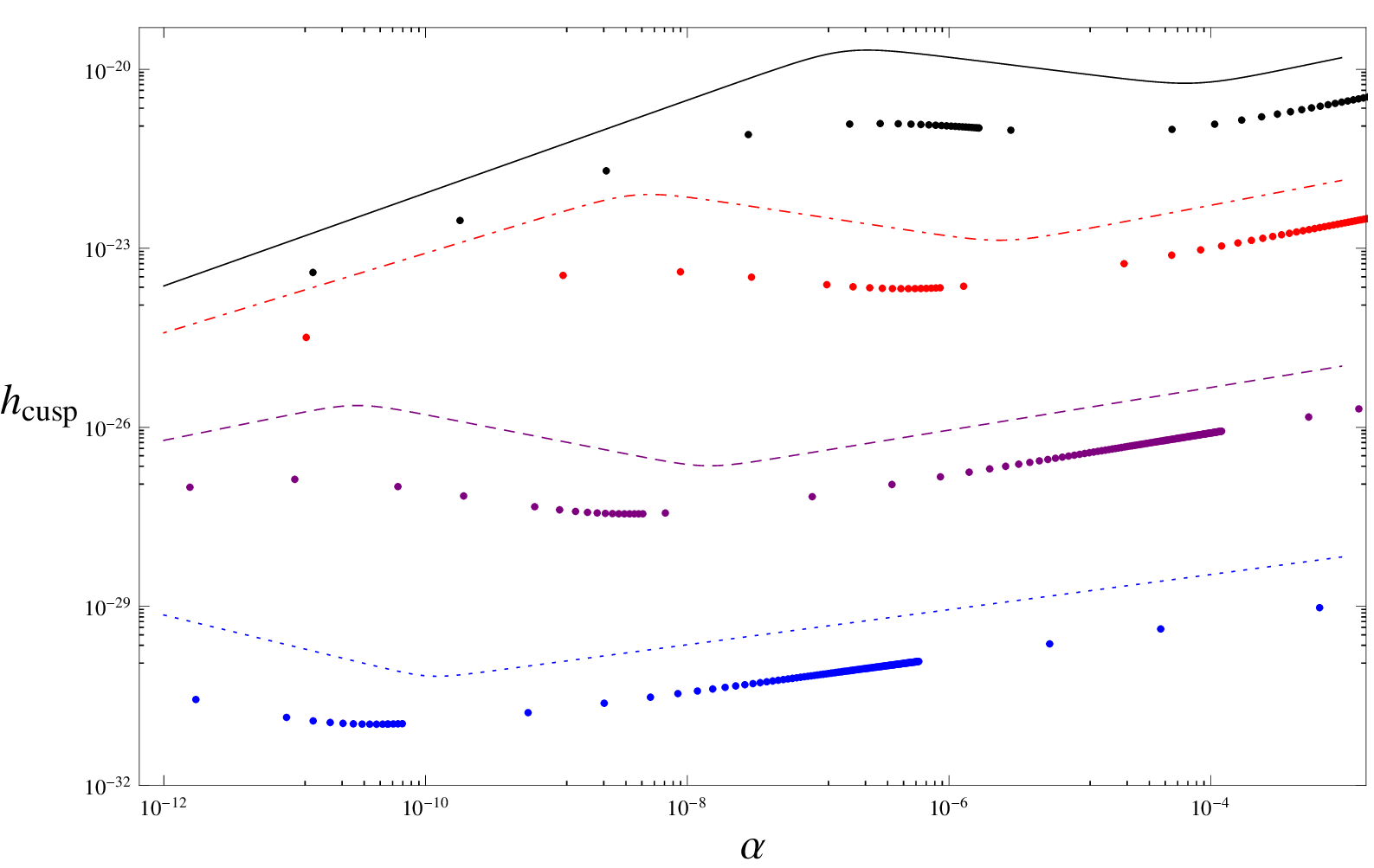}
\caption{
A log log plot of the  GW amplitude of bursts as a function
of $\alpha$ for a fiducial frequency $f=150$Hz, and a detection rate
of 1 per year. The lines (solid or dotted/dashed) represent the 
graphs obtained using the DV interpolating functions, allowing for 
a direct comparison with \cite{DV2}. The sets of individual
dots correspond to the exact numerical redshift integrations, where
we used the exact functions $t(z)$, $D_A(z)$, for the concordance
cosmology. The plots are colour coded, from the black, DV result
at the top, through red (dot-dash) for $n=1$, purple (dashed) for
$n=3$, and blue (dotted) for $n=6$ and all have an intercommutation
probability of $P=10^{-3}$.
} }

\FIGURE{\label{fig:Rate21}
\includegraphics[width=14cm]{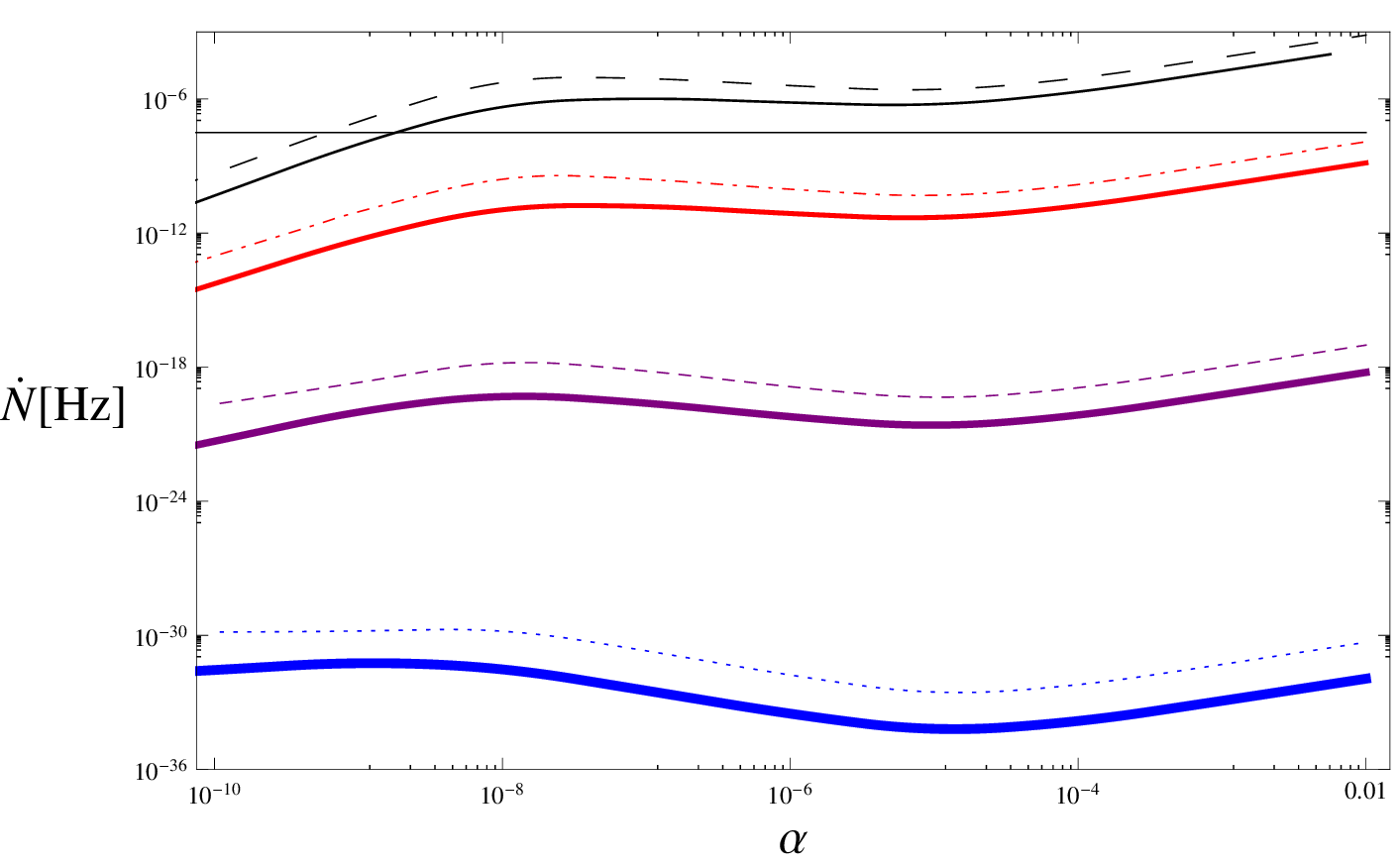}
\caption{
A similar plot to that of figure \ref{fig:All}, but in this case showing
the expected rate at an amplitude of $10^{-21} s^{-1/3}$, using the
method of \cite{Siem}. The plots are colour coded as in figure \ref{fig:All}
except in this case the thick solid lines are now the numerical 
integration results. From top to bottom: the 3D $P=10^{-3}$ result,
and the extra dimension plots with $n=1$, $n=3$, and $n=6$ respectively.
The horizontal black line indicates a rate of one event per year.
} }

An alternative way of presenting the GWB information is to instead
compute the expected detection rate of events with amplitudes greater
than (or equal to) a given amplitude. As explained in the papers of
Siemens et al.\ \cite{Siem, SiemStoch}, the one scale model used by DV, in 
which all loops are taken to have essentially the same length (\ref{onescale}), 
does not capture the full dynamical range of the cosmic string network
which will have loops, in theory, at all scales. They therefore recomputed
the rate in order to take into account the dependence not only on
redshift, but also on length scales, which they encoded in the
amplitude of the cusp waveform $A$, found by considering
\be
h^{cusp}=A |f|^{-1/3}
\ee
(in the logarithmic representation of DV) and comparing with (\ref{4Dwform}),
or in the extra dimensional case, (\ref{nDwform}),
resulting in a rate per redshift interval $dz$ and amplitude 
interval $dA$, rather than the rate per redshift interval obtained by DV. 
However, on using the one scale model (\ref{onescale}), where amplitudes
are directly associated with redshifts, the amplitude dependence is
effectively integrated out and the generalised expression found 
in \cite{Siem} reduces to (\ref{dndot}), the expression used by DV and 
hence to (\ref{newrate}) in the extra dimensional case.
The results in \cite{Siem} are therefore presented as a rate
plot against the string parameter $G\mu$, calculated by 
integrating $d\dot{N}$ out to redshift values corresponding to the 
chosen amplitude (we use their value of $A=10^{-21}$s$^{-1/3}$ only) 
at various values of $G\mu$:
\be
\frac{\varphi_t^{2/3}(z)}{(1+z)^{1/3}\varphi_r(z)}
=\frac{50A H_0^{-1/3}}{\alpha^{5/3}}
\ee
where $\alpha\sim50G\mu$ as usual, and $\varphi_t$, $\varphi_r$ are either 
the DV interpolating functions, or are related to the exact 
functions $t(z), D_A(z)$ (c.f.\ Siemens Appendix A \cite{Siem}). This equation 
relating the redshift and amplitude is 
easily found from the expression for the cusp waveform derived in \cite{DV}.
Figure \ref{fig:Rate21} shows the rates calculated for the 
3D and extra dimensional cases using this alternative approach.

\section{Discussion}
\label{disc}

The clear message of our results is that the impact of motion in 
extra dimensions can be significant. That extra dimensions should
have an effect is not unreasonable, since they can be viewed as
additional degrees of freedom living on the string\footnote{We thank
Jose Blanco-Pillado for discussions on this point.},
for example, superconducting cosmic strings, \cite{Wit}, can be
represented as a dimensional reduction of standard five dimensional
KK theory \cite{SCS}. For these cosmic strings, the currents round off 
cusps in much the same way as we have described here, \cite{scusp},
and also alter the balance between the energy and tension of the string,
\cite{SCgrav}, which has a clear gravitational impact. This naturally 
raises the concern that we may not be able to distinguish between
extra fields living on a cosmic string, and extra dimensions in which
the string is moving. This would certainly be the case if one was 
observing a string and a single GWB. However, our calculation was
for the expected signal from a cosmological network of strings, which 
depends not only on the GWB waveform, but also on the properties of 
the network.
Superconducting string networks have not been as well explored as
those of standard cosmic strings,
\cite{scnet}, with the main focus being on the different physics
induced by the long range electromagnetic interactions. Nevertheless,
as superconducting strings have similar intercommutation 
properties to standard cosmic strings \cite{sci}, it is likely
that the network is more similar to the usual cosmic string network
than that of the cosmic superstrings. Thus, while the individual 
GWB waveforms will be similar, the expected rates and signals we 
have calculated for the cosmological networks
are indeed specific to extra dimensions.

It is perhaps a little surprising that the effect of extra dimensions
can be so dramatic. We therefore now examine our assumptions carefully,
raising below a series of critiques together with a discussion of
their validity and resolution.

The basic reason for the suppression of the signal is the distribution
over the near cusp parameter $\Delta$. Our simulations with test loop
trajectories were performed in flat space with a toroidal (flat) 
compactification. Clearly the cosmological evolution will influence
the distribution of momentum across the various modes (although paradoxically
it would seem to damp higher momenta in the noncompact dimensions).
Since these strings form from the collision of a brane and anti-brane,
it seems likely that they have significant initial momentum in the
extra dimensions, thus we see no reason to curtail our solution space
in this way. 
The main objection to having total freedom of the internal modes is 
that by wrapping back and forth across the extra dimension(s) the string
has more opportunity to self intersect, thus curtailing the additional
freedom in that direction.

We modelled this effect by exploring the self-intersection of
a $1-3/1$ family of loops. We chose this combination of harmonics,
as the 4D family will have a 3D limit which were the first simple
loop trajectories explored by Kibble and Turok \cite{KT}. 
In 3D, the loop family self intersects approximately $30\%$ of the time
(note, the original plot of Turok \cite{KT} is inaccurate, see \cite{CDH}
for the correct version). When an additional dimension is introduced, 
the measure of the solution set allowing for self
intersections again becomes zero: the argument is once more parametric. 
A string will self-intersect if
\be
\Ab(\tau-\sigma) + \Bb(\tau+\sigma) = \Ab(\tau-\sigma') + \Bb(\tau+\sigma')
\label{selfint}
\ee
for some $\tau, \sigma, \sigma'$. Thus, there are three dynamical
variables and $3+n$ constraints. In 3D, it is therefore possible
to satisfy these constraints simultaneously, although a more
careful check of the parameters shows that not all loop solutions
can satisfy these constraints. Nonetheless, it shows how the subspace
of self intersecting loops can be of nonzero measure in parameter space.
With extra dimensions however, satisfying (\ref{selfint}) necessarily
requires a constraint on parameter space, hence the subspace with
self intersections will be of lower dimension than parameter space.
Even compactifying the extra dimensions does not change this argument,
unless we take into account the finite width of the string. 
Essentially, if we take the string to have zero width, then it can 
easily miss itself even when winding back and forth across the extra 
dimensions many times. However, with finite width, the self intersection
probability now becomes nonzero, and of order ${\cal O}(w/R)^n$ (where
$w$ is the string width, and $R$ the size of the extra dimension).
This therefore suggests that we introduce this ratio in a finite
width correction to the GWB measure.

\vskip 3mm

As we mentioned during our initial discussion, warping of the 
extra dimensions could also provide some significant dynamical
effect. The results of Avgoustidis \cite{AA} indicate that warping
is not as dramatic a trapping force as intially suspected, however, 
any confinement of strings could be significant, and a detailed modelling
of this effect is necessary. For now, we model a restriction in the extra
dimensional motion by a restriction in $\Delta$. Instead of allowing 
$\Delta$ to range over the full unit interval, we restrict 
$\Delta\in[0,\Delta_0]$. Thus we must renormalize ${\cal C}$:
\be
{\cal N}(\Delta_0) = \int_0^{\Delta_0} {\cal C}(\Delta) = 1 \qquad
\Rightarrow \qquad {\cal C}(\Delta) = \frac{n}{\Delta_0^n} \Delta^{n-1} \;.
\label{delta0C}
\ee
Thus the relevant $\Delta$ integral (\ref{deltaint}) now becomes
\be
\label{deltaint0}
\int_0^{{\rm min}\{\Delta_0,\theta_m\}} 
{\cal C}(\Delta) \left ( \theta_m(z) - \Delta \right )^2
= \theta_m(z)^2 {\cal F}_n \left [ \frac{\theta_m}{\Delta_0} \right ]
\ee
where
\be
{\cal F}_n [x] =
\frac{2 x^n}{(n+1)(n+2)} H[1 - x]
+ \left (1 - \frac{2n}{(n+1)x} 
+ \frac{n}{(n+2) x^2} \right ) H[x-1] \;.
\ee

A reasonable value for $\Delta_0$ might be to use the one parameter
ratio that does impact on the loop families and motion: the ratio
of string width to the size of the extra dimension. The Nambu action
is only a good approximation when the width of the string is small
compared to scales of physical interest. This is rarely a problem
in cosmology, as the string width is set by the inflationary scale,
and the size of the universe rapidly becomes large. As far as
the extra dimensions are concerned however, these are stabilized
at a couple of orders of magnitude above the string scale, hence
while the Nambu action is a good approximation, we might expect some
corrections to show up due to parameter restriction from self intersection
or excessive winding as already discussed. We therefore expect this
parameter to be related to the probability of intercommutation, which
can be viewed as arising because of the strings ``missing'' each other
in the internal dimensions. 

To test this alternate expression, we took
values of $\Delta_0 = 0.1, 0.01, 0.001$, and $10^{-4}$.
These values were motivated by a limiting sensible ratio $w/R$,
and the value of $P$. From (\ref{deltaint0}), we see that the
effect of $\Delta_0$ is to cut off the integral as $\theta_m$ 
grows. For $\theta_m < \Delta_0$, the dependence of the rate on
$\theta_m(z)$ remains that of the previous section, however, 
as $\theta_m$ grows, the functional dependence shifts towards
the $\theta_m(z)^2$ form of the 3D result. From the expression
for $\theta_m(z)$, (\ref{thetaofz}), we see that this is proportional
to $(G\mu)^{-1/3}$, hence the rates converge to the 3D value
sooner for smaller $G\mu$. 
\FIGURE{\label{fig:RateDel}
\includegraphics[width=14cm]{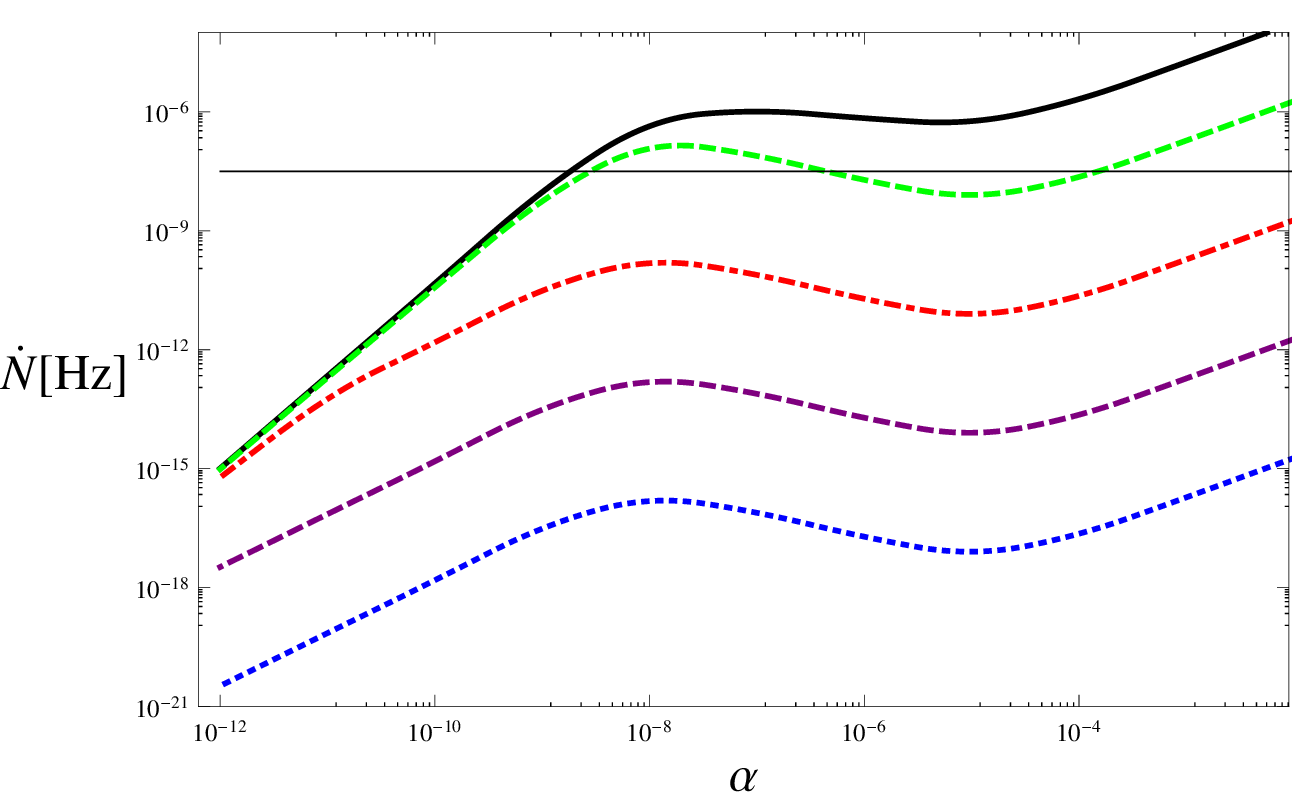}
\caption{
A plot of the dependence of the rate on $\Delta_0$ with the number
of extra dimensions fixed at three. From top to bottom, the 
solid black line indicates the 3D result, $\Delta_0 = 10^{-4}$ in 
dashed green, $\Delta_0 = 10^{-3}$ in dot-dash red, $\Delta_0 = 10^{-2}$
in dashed purple, and $\Delta_0 = 10^{-1}$ in dotted blue respectively.
The horizontal black line indicates a rate of one event per year.
} }

Figure \ref{fig:RateDel} shows the effect of the $\Delta_0$ parameter
on the event rate at an amplitude cutoff of $10^{-21} s^{-1/3}$ for
$\Delta_0$ ranging from $10^{-1} - 10^{-4}$ as indicated. Note that
once we use this more complicated expression (\ref{deltaint0}), the
use of the interpolating function approximation becomes too unwieldy,
and the rates had to be calculated by direct integration.
Figure \ref{fig:Raten} shows the effect of the rate on the number
of extra dimensions, fixing $\Delta_0 = 10^{-3}$ and allowing $n$
to vary as indicated. Here we see that for all $n$ the plots
converge on the 3D result at $\alpha \sim 10^{-11}$ but for 
$\alpha \sim 10^{-8}$ for example, the rate drops by roughly
an order of magnitude per dimension. A positive detection therefore
could potentially tell us the number of extra dimensions!
\FIGURE{\label{fig:Raten}
\includegraphics[width=14cm]{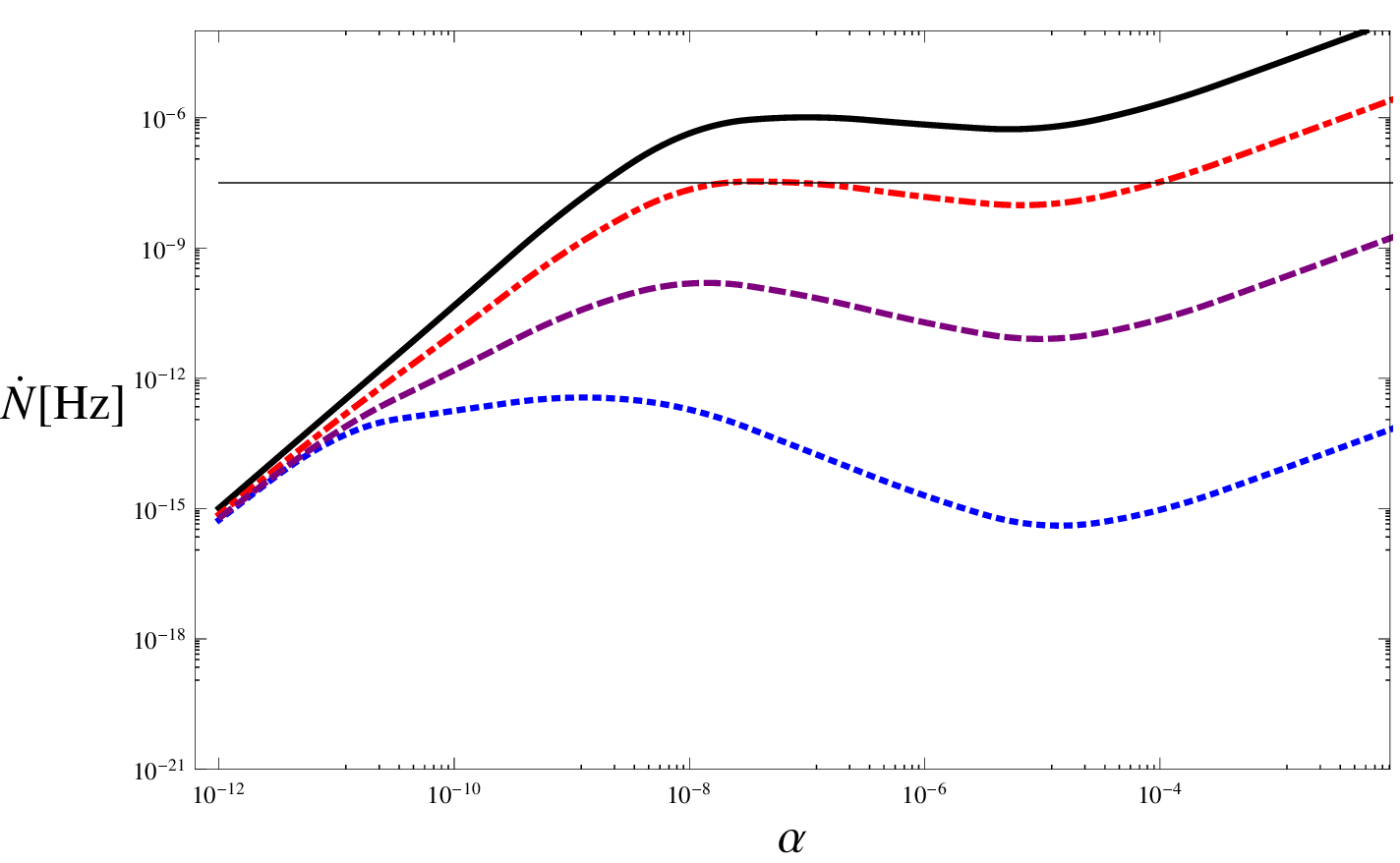}
\caption{
A recomputation of the rate plot with the new nearcusp measure 
(\ref{deltaint0}) fixing $\Delta_0 = 10^{-3}$, and allowing $n$ to vary.
From top to bottom, $n=0, 1, 3$ and $6$ respectively. 
The horizontal black line indicates a rate of one event per year.
} }

To sum up: We have included the effect of motion in extra dimensions
in the computation of the gravitational wave burst signal from cusp events
on cosmic string loops. We find a significant effect, even after
taking into account finite width effects and the size of the extra 
dimension. Clearly further work is required to get better control
of the approximations being used, in particular to take into account
more complex compactification geometries, however it does seem that 
motion in internal dimensions is important. Finally, if the
string tension lies in the serendipitous range $10^{-8}-10^{-10}$
then the possibility arises that a positive detection by gravitational
radiation would not only confirm the general brane inflation scenario,
but could provide a means of determining the number of (effective) 
extra dimensions.
 
\acknowledgments

We would like to thank A.\ Avgoustidis, J.J.\ Blanco-Pillado, D.\ Chung,
L.\ Leblond, S.\ Rajamanoharan, P.\ Shellard and G.Shiu for helpful discussions. 

RG and IZ would like to thank the Perimeter Institute for hospitality
while this work was being undertaken, and RG would like to thank the
Cambridge Centre for Theoretical Cosmology for hospitality while this
work was being completed. 

SC is supported by EPSRC, EOC is supported by EC FP6
through the Marie Curie EST project MEST-CT-2005-021074,
GG is suppported by the Government of Canada through Industry Canada and
by the Province of Ontario through the Ministry of Research, RG is
partially supported by STFC, and IZ is partially supported by the 
SFB-Transregio TR33 ``The Dark Universe" (DFG) and the EU FP7  
program PITN-GA-2009-237920.

\appendix
\section{Finite width corrections to the cusp}
\label{AppA}

In order to examine the validity of the Nambu approximation, 
we investigate the behaviour of the extrinsic curvature in 
the vicinity of an exact cusp. 
A 2D worldsheet living in 4 dimensions has codimension 2 and thus 
there exist two families of normals $n^\mu_i$, satisfying:
\bea
n^\mu_i n^\nu_j\eta_{\mu\nu}&=&-\delta_{ij}\\
n^\mu_i \frac{\partial X^\nu}{\partial\sigma_A}\eta_{\mu\nu}&=&0\;.
\label{normalcond2}
\eea
We choose our normals to be 
\bea
n_1^\mu&=&(1-({\bf a}^\prime.{\bf b}^\prime)^2)^{-\frac{1}{2}}\bigl
(1+{\bf a}^\prime\ldotp {\bf b}^\prime, \ab'+{\bf b}^\prime \bigr)\\
n_2^\mu&=&(1-({\bf a}^\prime.{\bf b}^\prime)^2)^{-\frac{1}{2}}\bigl
(0, {\bf a}^\prime \wedge {\bf b}^\prime \bigr )
\eea
both of which satisfy the above conditions. Even though $n^\mu_1$ and $n^\mu_2$ 
are not well defined at the cusp, they are well behaved at any distance 
or time arbitrarily close to it. As we approach the cusp, (which we take 
to be at $\tau=0$, $\sigma=0$ as in section \ref{gravitywaves3d}),  
$n^\mu_1$ becomes parallel to $\dot{X}$ and tilts over towards the light 
cone whereas $n_2^\mu$ flips direction across the cusp.

The extrinsic curvatures associated with $n^\mu_1$ and $n^\mu_2$ are:
\bea\label{ExCurvdef}
K_{_i AB}=\frac{\partial X^\mu}{\partial \sigma^A}\frac{\partial X^\nu}
{\partial\sigma^B}\nabla_{(\mu}n_{_i \nu)}=-n_{_i \mu }X^\mu_{,AB}~.
\eea
Notice that
\be
X^\prime=\Big(0, \frac{{\bf b^\prime-\ab^\prime}}{2}\Big),\;\;
\dot{X}=\Big(1,\frac{{\bf a^\prime+b^\prime}}{2}\Big)
\ee
\be
\Rightarrow \qquad
X^{\prime\prime}=\ddot{X}=\Big(0, \frac{{\bf a^{\prime\prime}
+b^{\prime\prime}}}{2}\Big),\;\;
\dot{X}^\prime=\Big(0,\frac{{\bf b^{\prime\prime}-\ab^{\prime\prime}}}{2}\Big)
\ee
and hence we can write all second derivatives as:
\be
X^\mu_{,AB}=\frac{1}{2}\bigl(0,
(\delta^\tau_A\delta^\tau_B+\delta^\sigma_A\delta^\sigma_B)({\bf
a^{\prime\prime}+b^{\prime\prime}})+2\delta^\tau_{(A}\delta^\sigma_{B)}({\bf
b^{\prime\prime}-\ab^{\prime\prime}})\bigr ).
\ee
Substituting this into (\ref{ExCurvdef}) yields
\bea
K_{_1 A B}&=&\frac{1}{2}(1-(\ab'.\bb')^2)^{-\frac{1}{2}}
\bigl[(\delta^\tau_A\delta^\tau_B+\delta^\sigma_A\delta^\sigma_B)
(\ab'' . \bb' + \bb'' . \ab')
+2\delta^\tau_{(A}\delta^\sigma_{B)}({\bb^{\prime\prime}}.\ab^\prime
-\ab^{\prime\prime}.\bb')\bigr] \nonumber \\
K_{_2 A B}&=&\frac{1}{2}(1-({\bf a}^\prime.{\bf b}^\prime)^2)^{-\frac{1}{2}}
\bigl[(\delta^\tau_A\delta^\tau_B+\delta^\sigma_A\delta^\sigma_B)
({\bf a^{\prime\prime}+b^{\prime\prime})\cdot (a^\prime \wedge b^\prime)}\\
&& \hskip 2cm +\, 2\, \delta^\tau_{(A}\delta^\sigma_{B)}
(\bb'' - \ab'')\cdot (\ab' \wedge \bb')\bigr ]\nonumber.
\eea

The leading nontrivial finite width correction to the Nambu action
appears at fourth order in $w$, the width of the string, and
is given by \cite{EffAction,Anderson}
\be
S=- {\bf \mu}\int d^2\sigma \sqrt{\gamma}\;(1+ w^4 
[\alpha_2(\Sigma_1+\Sigma_2)^2 + 2\alpha_3(\Sigma_3^2 + \Sigma_1\Sigma_2)]\;.
\label{action}
\ee
The $\alpha_n$ are numerically calculated coefficients dependent on
the specific model for the vortex, and for the Nielsen-Olesen vortex 
were computed in \cite{EffAction} and found to be of order unity.
For example, in the supersymmetric Abelian Higgs vortex, $\alpha_2 =
-\alpha_3/2 \sim 3.36$. The $\Sigma_i$ are 
scalars on the worldsheet constructed from the extrinsic curvatures:
\be
\Sigma_1=K_{_1 A B}K_1^{ A B},\;\;
\Sigma_2=K_{_2 A B}K_2^{ A B},\;\;
\Sigma_3=K_{_1 A B}K_2^{ A B}.
\ee
Note that to calculate these scalars we raise the indices of  $K_{_i AB}$ using
the inverse of our worldsheet metric (\ref{gammaAB}): 
\be
\gamma^{AB}=\frac{2}{(1-{\bf a}^\prime \cdot{\bf b}^\prime)}~\eta^{AB}\;.
\ee
We thus calculate the worldsheet scalars to be:
\bea
\Sigma_1&=&K_{_1 A B}K_1^{ A B}\nonumber\\&=&2(1-({\bf a}^\prime.{\bf
b}^\prime)^2)^{-1}(1-{\bf a}^\prime .{\bf b}^\prime)^{-2}[({\bf
a^{\prime\prime}\cdot b^\prime+b^{\prime\prime}\cdot
a^\prime})^2-(\bf b^{\prime\prime}}\cdot \ab^\prime-{\bf a^{\prime\prime}\cdot
b^\prime)^2]\nonumber
\\&=&\frac{8~{\bf
(a^{\prime\prime}\cdot b^\prime)( b^{\prime\prime}\cdot
a^\prime})}{(1-({\bf a}^\prime.{\bf b}^\prime)^2)(1-{\bf
a}^\prime .{\bf b}^\prime)^{2}}
\\ \nonumber \\
\Sigma_2&=&K_{_2 A B}K_2^{ A B}\nonumber
\\&=&2(1-({\bf a}^\prime.{\bf
b}^\prime)^2)^{-1}(1-{\bf a}^\prime .{\bf b}^\prime)^{-2}
[\bigl(({\bf a^{\prime\prime}+b^{\prime\prime})\cdot (a^\prime \wedge
b^\prime)}\bigr)^2 -\bigl (({\bf
b^{\prime\prime}-\ab^{\prime\prime})\cdot (a^\prime \wedge
b^\prime)}\bigr )^2]\nonumber \\&=&\frac{8~{\bf ({\bf
a^{\prime\prime}\cdot ( a^\prime \wedge b^\prime)}) ({\bf
b^{\prime\prime}\cdot (a^\prime \wedge b^\prime)}})}{(1-({\bf
a}^\prime.{\bf b}^\prime)^2)(1-{\bf a}^\prime .{\bf b}^\prime)^{2}}
\\ \nonumber \\
 \Sigma_3&=&K_{_1 A B}K_2^{ A B}\nonumber
\\&=&2(1-({\bf a}^\prime.{\bf
b}^\prime)^2)^{-1}(1-{\bf a}^\prime .{\bf b}^\prime)^{-2}[({\bf
a^{\prime\prime}\cdot b^\prime+b^{\prime\prime}\cdot
a^\prime})\bigl(({\bf a^{\prime\prime}+b^{\prime\prime})\cdot
(a^\prime \wedge b^\prime)}\bigr) \nonumber
\\&&-(\bb^{\prime\prime}}\cdot \ab^\prime-{\bf a^{\prime\prime}\cdot
b^\prime)\bigl ((\bb^{\prime\prime}{\bf
-a^{\prime\prime})\cdot (a^\prime \wedge
b^\prime)}\bigr )]\nonumber \\&=&\frac{4~[{\bf(a^{\prime\prime}\cdot
b^\prime) (b^{\prime\prime}\cdot(a^\prime \wedge b^\prime)})+
{\bf(b^{\prime\prime}\cdot a^{\prime})( a^{\prime\prime}\cdot
(a^\prime \wedge b^\prime)})]}{(1-({\bf a}^\prime.{\bf
b}^\prime)^2)(1-{\bf a}^\prime .{\bf b}^\prime)^{2}}\;.
\eea
In order to determine if these terms will result in a significant
correction to the action (\ref{action}), we must examine the scalars' 
behaviour close to our exact cusp. We perform a Taylor expansion 
of $\ab'$, $\bb'$, 
$\ab''$ and $\bb''$ around the cusp position $\sigma_\pm=0$, yielding
\bea
\Sigma_1&\simeq&32\biggl [\frac{ {\rm (a_0''^2\sigma_--
a_0''b_0''\cos\psi~\sigma_+)(b_0''^2\sigma_+-a_0''b_0''\cos \psi~ \sigma_-)}}
{(a_0''^2\sigma_-^2+b_0''^2\sigma_+^2-2a_0''b_0''\cos\psi~\sigma_-\sigma_+)^3}
\biggr ]\nonumber \\
\\
\Sigma_2&\simeq&32\biggl [\frac{ {\rm a_0''^2 b_0''^2 
\sin^2\psi~\sigma_-\sigma_+} }
{(a_0''^2\sigma_-^2+b_0''^2\sigma_+^2-2a_0''b_0''\cos\psi~\sigma_-\sigma_+)^3}
\biggr ]\nonumber \\
\\\Sigma_3&\simeq&-16\biggl [\frac{{\rm a_0''b_0''\sin \psi }}{
(a_0''^2\sigma_-^2+b_0''^2\sigma_+^2-2a_0''b_0''\cos\psi~\sigma_-\sigma_+)^2}
\biggr ]~,\nonumber \\
\eea
where $\psi$ is the angle between ${\bf a_0''}$ and ${\bf b_0''}$
and we use ${\rm a_0''}$ and ${\rm b_0''}$ to refer to the magnitude of 
their corresponding vectors. We now fix $\sigma=0$ and allow $\tau$ 
to vary in order to see how the string curvature behaves as the cusp 
forms:
\be
\label{sigmanearcusp1}
\Sigma_1\simeq32\biggl [\frac{ {\rm (a_0''^2-
a_0''b_0''\cos\psi)(b_0''^2-a_0''b_0''\cos \psi)}}{
(a_0''^2+b_0''^2-2a_0''b_0''\cos\psi)^3}\biggr ]\tau^{-4},~~~~~~\sigma=0, 
~ \tau\ll L\;.
\ee
Alternatively, we can investigate the shape of the string at the precise 
moment of the cusp by keeping $\tau=0$ and allowing $\sigma$ to vary:
\be
\Sigma_1\simeq-32\biggl [\frac{{\rm (a_0''^2+a_0''b_0''\cos\psi)
(b_0''^2+a_0''b_0''\cos \psi)}}{(a_0''^2+b_0''^2+2a_0''b_0''\cos\psi)^3}\biggr]
\sigma^{-4},~~~~~~\tau=0,~\sigma\ll L\;.
\label{sigmanearcusp2}
\ee

We can see that as the cusp is formed ($\tau\to 0$), the extrinsic curvature
scalars $\Sigma_i$ grow rapidly, diverging as $\sigma_{\pm}^{-4}$ and becoming 
infinite at the cusp itself ($\Sigma_2$ and $\Sigma_3$ exhibit similar 
behaviour to $\Sigma_1$). These terms can therefore no longer be 
neglected and it would seem that we must indeed include the leading 
order correction in (\ref{action}). 

However, we note that, as in section \ref{gravitywaves3d}, we can take the 
order of magnitude of  $a''_0 \sim b''_0 \sim 2\pi / L$ (see \cite{DV} also), 
implying
\be
\label{SigmaL}
\Sigma_i\sim \frac{L^2}{\sigma_\pm^4}\;.
\ee
Thus the corrections to the Nambu action become
non-negligible for
\be
\Sigma_i \sim \frac{1}{w^2}
\ee
i.e.
\be 
\label{sigmamin}
\sigma_\pm \sim (w L)^{1/2}. 
\ee
We therefore conclude that the Nambu action (\ref{NGaction}) will not
break down and our analysis remains valid provided we take 
\be 
\sigma_\pm \gg (w L)^{1/2}. 
\ee
As we mentioned in section \ref{disc}, the width of the string $w$ 
is set by the inflationary scale, while the physical scale of 
interest is cosmological, and therefore $(wL)^{1/2}$ will 
be extremely small, making it possible to fulfil this condition.

The idea is therefore that if an exact cusp is supposed to 
occur at $\sigma_\pm=0$, where the $\ab'$ and $\bb'$ curves cross, 
then at some point before the event (e.g. at $\tau\sim (wL)^{1/2}$ 
and $\sigma=0$), the condition $\sigma_\pm> \sigma_{min} \sim (wL)^{1/2}$ 
is broken and the Nambu approximation breaks down at that moment. 
It has been argued that as the cusp forms, the two string segments 
close to the point of the cusp could overlap (c.f.\ figure 1 
of \cite{CuspOverlap2}), resulting in a small loop separating 
from the string (due to it reconnecting), along with the consequent 
particle emissions. This would then result in a bridging effect 
and a rerouting of the trajectories on the Kibble-Turok sphere 
\cite{CuspOverlap, CuspOverlap2}. Given that Olum and 
Blanco-Pillado also show that the size of the overlapping segment is 
of the same order as the value of $\sigma_\pm$ in (\ref{sigmamin}), 
(i.e.\ where we believe the curvature effects are becoming relevant), it 
would seem that using the analytical description found from the Nambu action 
to calculate  the overlap at this point is unjustified. However, 
the general picture of an additional emission of energy of the 
same order as the energy in that segment of the string,  
consistent with the simulation in \cite{CuspOverlap2},  still 
seems feasible since we expect that as the curvature starts to diverge, 
the string must somehow round off to avoid an exact cusp.

We now use a simple method of estimation to check if GWB's 
from a cusp will be affected by the imposition of a lower bound 
of $\sigma_{min} \sim (wL)^{1/2}$, allowing us to continue using 
the Nambu approximation. If we consider eqn (\ref{Iplusminus}) and 
set $\varepsilon=0$, (i.e.\ $k^{\mu} \parallel \ell^{\mu}$), the 
integrals $I_\pm$ can be estimated by
\bea \label{approxintu}
I=\int^\infty_{-\infty} du~ u~ e^{\pm i u^3}
\eea
where $u$ is given by (\ref{IntReparam}) as before. However the 
real limits of this integral should be determined by the limits 
on $\sigma_\pm$. In other words (\ref{approxintu}) is an approximation of 
\bea \label{exactintu}
I_{\sigma}=\int^{u_{max}}_{-u_{max}} du~ u~ e^{\pm i u^3}-
\int^{u_{min}}_{-u_{min}} du~ u~ e^{\pm i u^3}
\eea 
where $u_{min}\equiv u(\sigma_{min})$ and $u_{max}\equiv u(L)$. We 
illustrate how these limits influence the integral's 
behaviour in figure \ref{Uintegral}.
\FIGURE{  
\includegraphics[width=10cm]{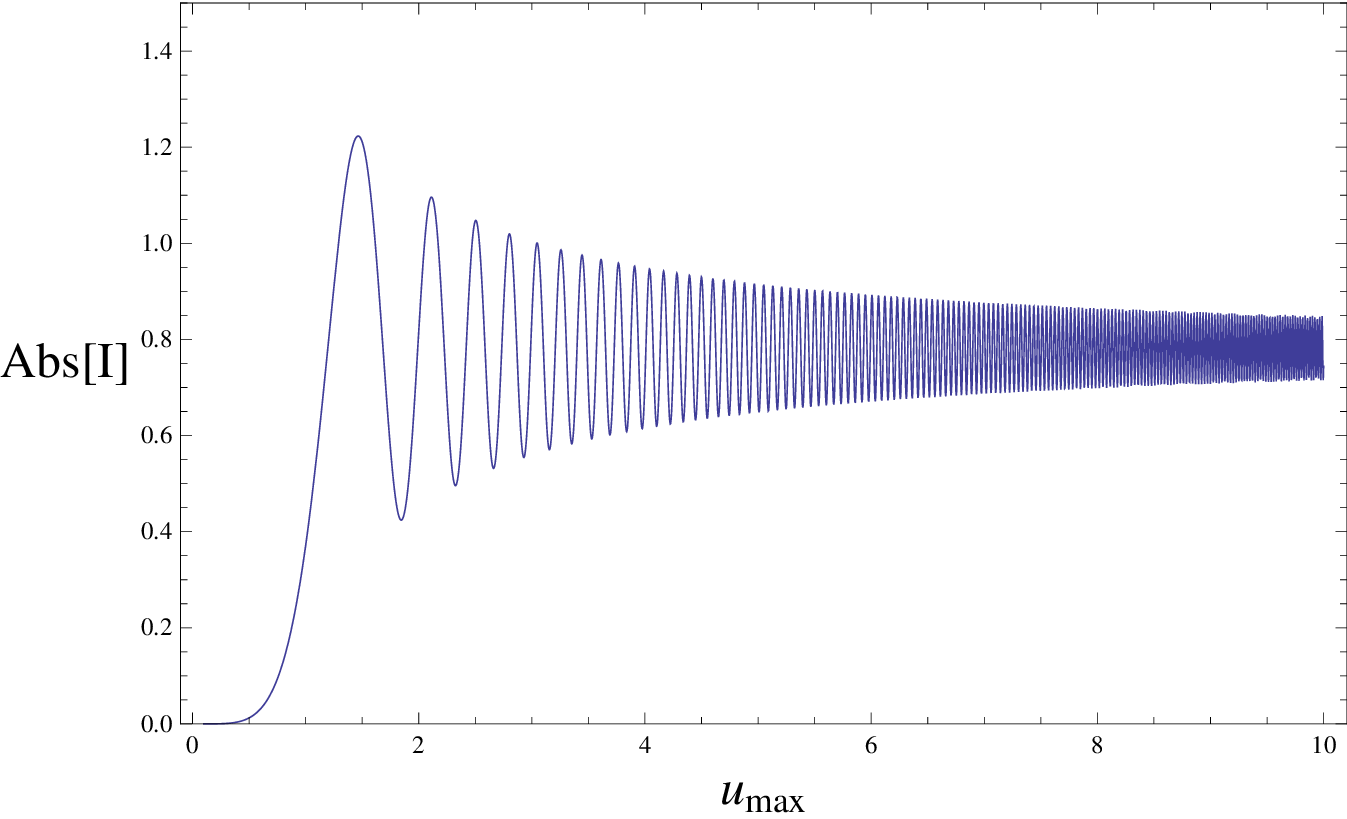}
\caption{The value of the integral $I$ (\ref{approxintu}) integrated 
between $\pm u_{max}$ for various values of $u_{max}$.}
\label{Uintegral}
}
We can see from this plot that the approximate value of $0.7818$ obtained 
from (\ref{approxintu}) is roughly within $20\%$ of the exact 
value of $I_{\sigma}$ if we assume that $u_{min}\lesssim0.5$ 
and $u_{max}\gtrsim4$.
We now proceed to calculate the value of $u_{min}$ using our 
lower bound of $\sigma_{min}\sim(wL)^{1/2}$:
\bea
u_{min}&\sim& \left[\frac{1}{12}|m|\omega_L (\ddot{X}_\pm)^2
\right]^{\frac{1}{3}}\cdot \sigma_{min}\nonumber\\
&\sim& \left[\frac{1}{12} \cdot 2\pi f\cdot 
\Big(\frac{x_2}{L}\Big)^2\right]^{\frac{1}{3}} \cdot \sigma_{min}\nonumber\\ 
&\sim&  \left[\frac{\pi}{6}f \Big(\frac{x_2}{L}\Big)^2\right]^{\frac{1}{3}} 
\cdot (w L)^{1/2}
\eea
where we have substituted $f=m\omega_L/2\pi$ as before 
and $\ddot{X}_{\pm}\sim x_2/L$. We introduce $x_2$ as a 
naive but simple way of incorporating the number of harmonics 
in the string solution and as an attempt to account for how 
wiggly the string is. In the DV approximation it is taken to be 
of order one (we have used $\ddot{X}_{\pm}\sim 2\pi/L$ previously). 
We again use the one scale model (\ref{onescale}) 
with $\alpha\sim \Gamma G \mu$ and $\Gamma\sim 50$ and the DV 
interpolating function $t\sim t_0(1+z)^{-3/2}(1+z/z_{eq})^{-1/2}$ \cite{DV}. 
Now if we use our fiducial frequency $f=150$ Hz $\sim 10^{-22}$ GeV, 
which is in the optimal frequency range of LIGO, and $10^{-7} > 
G\mu> 10^{-12}$, $t_0^{-1}\sim H_0\sim 10^{-42}$ GeV, $G\sim 10^{-38}$ 
GeV$^{-2}$  and  $w\sim \frac{1}{\sqrt{\mu}}$, we get:
\bea
u_{min}&\sim& f^{1/3}x_2^{2/3}(\Gamma G\mu)^{-1/6}t^{1/6}\mu^{1/4} \nonumber\\
&\sim& 10^{-7} x_2^{2/3} \Gamma^{-1/6}(G\mu)^{-1/6} 
\mu^{1/4} H_0^{1/6}(1+z)^{1/4}\Big(1+\frac{z}{z_{eq}}\Big)^{1/12} 
\nonumber\\&\sim&10^{-14}10^{-9.5} x_2^{2/3} \Gamma^{-1/6} 
(G\mu)^{-5/12}(1+z)^{1/4}\Big(1+\frac{z}{z_{eq}}\Big)^{1/12}\nonumber\\
&\sim&10^{-18} x_2^{2/3} \Big(\frac{f_{Hz}}{150}\Big)^{1/3} 
\Big(\frac{\Gamma}{50}\Big)^{-1/6} \Big(\frac{G\mu}{10^{-12}}
\Big)^{-5/12}(1+z)^{1/4}\Big(1+\frac{z}{z_{eq}}\Big)^{1/12}
\eea
We see that as long as $x_2$ is not drastically large (i.e.\ the 
string is extremely wiggly)  or $\Gamma$ is not significantly 
smaller than $50$, $u_{min}\ll1$ is a valid approximation  
for any feasible gravitational wave experiment and none of the other 
parameters in the above expression can compensate for the smallness 
of the size of the cusp segment compared to the size of the loop. 
Indeed, even for very high redshifts up to $z\sim 10^{12}$ corresponding 
to $100$ MeV scales, the corrections at most change the result 
of $u_{min}$ by a few orders of magnitude.

We perform a similar calculation to approximate $u_{max}$ 
using $\sigma_{max}=L$:
\bea
u_{max}&\sim&  \left[\frac{\pi}{6}f \Big(\frac{x_2}{L}\Big)^2
\right]^\frac{1}{3} \cdot L\nonumber\\
&\sim&10^3~  x_2^{2/3} \Big(\frac{f_{Hz}}{150}\Big)^{1/3} 
\Big(\frac{\Gamma}{50}\Big)^{1/3} \Big(\frac{G\mu}{10^{-12}}
\Big)^{1/3}(1+z)^{-1/2}\Big(1+\frac{z}{z_{eq}}\Big)^{-1/6}
\eea
which suggests that the approximation remains valid at least up 
to $z\leq 10^8$ or $10^{-2}$ MeV scales. Hence (\ref{approxintu}) is a 
valid approximation of (\ref{exactintu}) for our limits on $\sigma_{\pm}$.

We have therefore shown that the use of the Nambu action (\ref{NGaction}), 
rather than the corrected action (\ref{action}), in calculating GWB's 
from an exact cusp is justified.

\end{document}